\documentclass[aoas,preprint]{imsart}

\RequirePackage[OT1]{fontenc}
\RequirePackage{amsthm,amsmath}
\RequirePackage[numbers]{natbib}

\usepackage{graphicx,psfrag,dcolumn,amsmath,amsthm,float,amssymb,color}
\usepackage{rotating}

\theoremstyle{remark}
\newtheorem{remark}{Remark}%[subsection]

\DeclareMathOperator{\Var}{\mathbb{V}ar}
\DeclareMathOperator{\Prob}{\mathbb{P}}
\DeclareMathOperator{\Expec}{\mathbb{E}}

\arxiv{math.PR/0000000}

\startlocaldefs
\numberwithin{equation}{section}
\theoremstyle{plain}

\endlocaldefs

\begin{document}

\begin{frontmatter}

\title{Scanning a Poisson Random Field for Local Signals}
\runtitle{Scanning a Poisson Random Field for Local Signals}

\begin{aug}
\author{\fnms{Nancy} \snm{R. Zhang}\thanksref{t3,t4,m3}\ead[label=e1]{nzh@wharton.upenn.edu}},
\author{\fnms{Benjamin} \snm{Yakir}\thanksref{t2,m2}\ead[label=e2]{msby@mscc.huji.ac.il}},
\author{\fnms{Charlie} \snm{L. Xia}\thanksref{t3,m4}\ead[label=e4]{li.xia@stanford.edu}}
\and
\author{\fnms{David} \snm{Siegmund}\thanksref{t2,m1} \ead[label=e3]{dos@stat.stanford.edu}}
\thankstext{t2}{supported by the National Science Foundation}
\thankstext{t3}{supported by NIH R01 HG006137-01}
\thankstext{t4}{supported by the Sloan Foundation}
\runauthor{Zhang, Yakir and Siegmund}

\affiliation{University of Pennsylvania \thanksmark{m3}, The Hebrew University of Jerusalem\thanksmark{m2}, and Stanford University \thanksmark{m1}}

\address{Department of Statistics\\The Wharton School, University of Pennsylvania\\Philadelphia 19004\\U.S.A.\\
\phantom{E-mail:\ }\printead{e1}}

\address{Department of Statistics\\The Hebrew University of Jerusalem\\Jerusalem 91905\\Israel\\
\phantom{E-mail:\ }\printead{e2}}
\address{Department of Medicine\\Stanford University School of Medicine\\ 269 Campus Drive\\Stanford, CA  94304\\U.S.A.\\
\phantom{E-mail:\ }\printead{e4}}

\address{Department of Statistics\\Stanford University\\ Sequoia Hall\\390 Serra Mall\\Stanford, CA  94305-4065\\U.S.A.\\
\phantom{E-mail:\ }\printead{e3}}

\end{aug}

\begin{abstract}
The detection of local genomic signals using high-throughput DNA sequencing data can be cast as a problem of scanning a Poisson random field for local changes in the rate of the process.  We propose a likelihood-based framework for for such scans, and derive formulas for false positive rate control and power calculations.  The framework can also accommodate mixtures of Poisson processes to deal with over-dispersion.  As a specific, detailed example, we consider the detection of insertions and deletions by paired-end DNA-sequencing.  We propose several statistics for this problem, compare their power under current experimental designs, and illustrate their application on an Illumina Platinum Genomes data set.
\end{abstract}

%\begin{keyword}[class=AMS]
%\kwd[Primary ]{60K35}
%\kwd{60K35}
%\kwd[; secondary ]{60K35}
%\end{keyword}


\end{frontmatter}

\section{Introduction}

Modern biology, especially genetics and fMRI analysis, has motivated a
great deal of theoretical and applied research in detection of local
signals in large fields of data.   See, for example, Lander and
Botstein (1989), Karlin, Dembo and Kawabata (1990), Feingold et al.
(1993) Worsley (1992), Siegmund and Worsley (1995).  Typically, the
random field representing the data is standardized, say, to have mean
zero when there is no signal and to have mean different from zero in
neighborhoods of signals.  These signal detection problems are statistically
irregular, since the parameters quantifying the magnitude and location of
each signal are confounded.

In many formulations the random field is assumed to be Gaussian, often
because of arguments based on the central limit theorem.
Control for the multiple comparisons involved in searching the field
for local signals is achieved by using the theory of maxima of
Gaussian fields to obtain a significance threshold
that controls the overall false positive rate.  This requires
that the normal distribution provide an adequate approximation
in the extreme tail of the distribution, which in turn
suggests that one be skeptical of the accuracy of the resulting thresholds,
especially in many cases where Poisson like data are involved and the
Poisson rate is not large.  Papers containing more precise asymptotic analyses
under various special conditions are
Rabinowitz and Siegmund (1995), Tang and Siegmund (2001), Peng
and Siegmund (2005), Chan and Zhang (2006), Siegmund, Yakir and Zhang (2011).

This paper is motivated by a number of problems arising from
high-throughput DNA sequencing data,
where the random field is assumed to be a
Poisson process, possibly non-homogeneous, or in some cases a mixture
of Poisson processes to deal with over-dispersion.  The signal to be
detected involves a local change in the rate of the Poisson process.
Such scans for local signals arise in the detection of DNA copy
number variations in DNA sequencing,  transcription
factor binding sites in chromatin immuno-precipitation followed by
sequencing (ChIP-Seq, see for example Schwarzman et al. 2013),
alternative transcription start and end sites in RNA sequencing, and
genomic insertions and deletions in paired-end DNA sequencing.  A brief
description of these motivating applications is given in the next section, but our focus will be on the last problem, also referred to as DNA structural variant detection.
Although these problems differ in biological context and model
formulation, they can be studied by closely related statistical methods.

We propose a framework for detecting local signals in Poisson-like data.  Scan statistics for the
applications mentioned in the preceding paragraph can be cast in the proposed framework.
We derive approximations for the false positive rates of likelihood ratio- and
score-based scan statistics for reasonably general Poisson random fields.
We also study the power of these statistics as a function of the
baseline rate and other parameters.  For the structural
variant detection problem, we first introduce, as illustration, a toy
mixture model that can be directly compared to some simple models for which
there is existing theory.  We then consider more complex models that
are more carefully tailored to the specific features of paired end reads
of DNA sequencing data.

This paper is organized as follows.  First, we describe some motivating applications in Section \ref{sec:examples}.  In Section \ref{sec:model}, we give a general framework for scans of Poisson random fields, first illustrating it on a simple mixture model (Section \ref{sec:simple}) and then on a more complex and realistic model for the problem of structural variant detection (Section \ref{sec:bettermodel}).  The simple mixture model has the benefit of being more transparent and allowing more direct comparisons to some existing scan statistics, while also lending qualitative insights that are transferable to more complex settings.  In Sections 4 and 5, we describe the procedure for p-value approximation for scan statistics on Poisson random fields. These approximations are derived for the simple mixture model in Section 6, with their accuracy examined by numerical experiments.  Even for the simple mixture model, it is not clear how to design the scan statistic with respect to the unknown parameters to maximize power under the different types of alternatives.  In Section \ref{sec:power}, we explore some of the complicated issues relating to power.   In Section 7, we return to the more realistic models for structural variant detection formulated in Section \ref{sec:bettermodel}.  Their p-value approximations are given, the power of various scan regimes under current experimental designs are studied, and and the analysis of a real data set from Illumina's Platinum Genomes is described. We conclude with a discussion in Section \ref{sec:discussion}.

The theory and methods described in this paper are at the core of \texttt{SWAN}, a comprehensive statistical pipeline for genomic structural variant detection.  \texttt{SWAN} is an open source R library available at:

\begin{center}\texttt{https://bitbucket.org/charade/swan/wiki/Home}.\end{center}

\section{Motivating Examples from Sequencing Experiments}\label{sec:examples}

High throughput short read sequencing, often referred to as
``next-generation sequencing,''  is a revolutionary way of quantifying DNA,
RNA, protein binding, and many other genome-wide features in biology.
The series of online supplementary articles of the November 2009 issue of Nature provide a good overview of the technology
and its applications.  As our main example, we consider DNA sequencing, which is described in the review by Medvedev, Stanciu and Brudno (2009).
Briefly, double-stranded DNA is extracted from the sample of interest and fragmented, followed
by the sequencing of a fixed number of bases, called reads, from one or
both ends of each fragment.  The lengths of the
fragments are selected to be within a specific range, e.g. 200 bases
with a standard deviation of 10.  When both ends of the fragment are sequenced, the data are referred to as paired ends, since the reads are paired, with one read coming
from each end of the double-stranded DNA molecule.  Since sequencing is unidirectional and proceeds only in the 5' to 3' direction of the DNA molecule, one read of each pair should come from the plus strand of each double stranded fragment, with the  other read of the pair coming from the minus strand.  The sequenced reads are then mapped to a reference genome, and the start positions, as well as orientation (plus or negative strand) of each read are recorded among many other features.  In obvious nomenclature, reads that map to the plus and negative strand of the template sequence are called, respectively, ``plus strand reads'' and ``minus strand reads''.  The mapped insert length is defined as the number of bases between the start position of the minus strand read and that of the plus strand read\footnote{Some papers define insert length to be the \emph{end} of the minus strand read minus the start of the plus strand read, which is our definition plus an read length.}.  If the sequenced genome is identical to the reference genome in the region spanned by the reads, then the mapped insert length is simply the length of the fragment from which the read pair derived, minus the length of one read, and also, each read pair should consist of one plus strand read and one negative strand read.  Important fixed quantities
in our ensuing models, which are chosen during the experiment, are the length of each read, $R$, and the distribution of the insert lengths,
which we characterize by a distribution function $F$ with mean $\delta$ and standard deviation $\sigma$.

\subsection{Detection of Changes in DNA Copy Number} \label{sec:cnv}
In diploid organisms, each cell has 2 copies of every chromosome.  A copy number change refers to the deletion or duplication of a chromosomal segment.  DNA sequencing
has been used to detect copy number change, because the density of
reads mapped to a genome interval depends on the relative quantity of that piece of DNA in the sequenced sample (Campbell et al., 2008; Chiang et al., 2009; Abyzov et al, 2011; Shen and Zhang, 2012).  Consider a simplified model for single-end sequencing, where the start positions of the mapped reads can be assumed to follow a non-homogeneous Poisson process $N(t)$ of intensity $\rho(t)$.  That is, for $s<t$, $N(t)-N(s)$ is the number of reads that map to the region $(s,t]$ on the reference genome.  We will call $N(t)$ the \emph{coverage process}. The function $\rho(t)$ depends not only on the copy number but also on other features, such as GC content, that are local to the neighborhood of $t$.  To control for local biases, Shen and Zhang (2012) considered a control process $M(t)$ derived from the sequencing of a control sample, and used sequential clustering of jumps in $N(t)-M(t)$ to detect copy number changes.  In a region free of CNV, the jumps form a symmetric random walk with increments 1 or - 1, while a clustering of jumps of one kind indicates the presence of CNV.  In this way the detection of CNV is
reduced to detection of an interval where a simple symmetric random walk shows an excess of 1's or -1's.  This problem has been well studied and will not be discussed here, except to note that knowledge of $\rho(t)$ is unnecessary, since information about the proximity in base pairs of one jump from another is ignored by this statistic.

Considerable effort has been made to estimate $\rho(t)$ using measurable
genomic features such as GC content and mappability (see, for example, Benjamini and Speed, 2012).  Assuming that a reliable estimate of $\rho(t)$ is available, we can model a deletion
of $(t_1, t_2]$ as a drop of the intensity function to
$\exp(\beta) \rho(t)$ within the interval, where $\beta < 0$.  The
parameter $\beta$ reflects whether the deletion is heterozygous or homozygous, and the purity of the deletion in the sample.
The log likelihood ratio for the process $N_t$, with and without the deletion is
\begin{equation} \label{deletion}
\beta [N_{t_2} - N_{t_1 - R}]  - [\exp(\beta) -1]
\int_{t_1-R}^{t_2}
\rho(t) dt.
\end{equation}
Since the boundaries of the deletion are unknown, a scan statistic would involve maximization of  (\ref{deletion}) over an appropriate  range of $t_1 < t_2$, and perhaps also over a reasonable range for $\beta.$  This model can obviously also be used to detect duplications.

\subsection{Detection of Structural Variants} \label{sec:sv}
Structural variants are insertions, deletions, inversions, and translocations of segments of DNA in the
genome.  Deletions result in a loss of copy number, and
insertions of DNA from another region of the genome result in a copy
number gain of that region.  Thus, structural variation sometimes cause copy number variation, but not always.  For example, a translocation, which is the
movement of a DNA segment from one position in the genome to
another, does not result in a change of copy number, but can be
viewed as a deletion at the original site followed by
an insertion at the new site.

\begin{figure}
\begin{center}
\includegraphics[scale=.6]{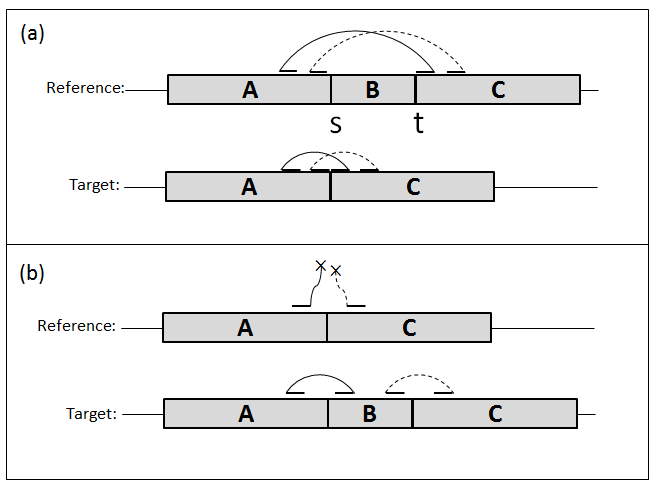} \\
\end{center}
\caption{\label{fig:schematic} Mapping of paired-end reads in region of deletion (a) and insertion (b).  In (a), the deleted region, labeled $B$, spans bases $s$ to $t$ in the reference genome, and is flanked by regions A and C.  Read pairs that span the deletion point, i.e. with the plus read in $A$ and minus read in $C$ in the target, map further apart in the reference than expected.  In (b), the inserted segment B is flanked by regions A and C in the target genome, with A and C joined together in the reference genome.  Fragments containing the A-B boundary produce hanging minus strand reads, while those that contain the B-C boundary produce hanging plus strand reads.  Deletions also produce hanging reads, if the end of a fragment overlaps with the boundary between $A$ and $C$ in target.  Similarly, for insertions, fragments that contain the entire inserted segment produce read pairs that map closer together than expected. }
\end{figure}

Structural variants are always parameterized with respect to the
reference genome template to which reads are mapped, and thus, for example,
deletion of $[s,t]$ refers to deletion of the DNA sequence starting at $s$ and ending at $t$ in the reference genome.  Paired-end DNA sequencing allows
the detection and sometimes the precise positioning of structural variants.
Figure \ref{fig:schematic} shows how deletions and insertions produce tell-tale patterns in the mapping of paired-end reads.  Figure \ref{fig:schematic} (a) shows a deletion of bases $s$ to $t$ in the reference genome.  The deleted region is labeled $B$ and is spanned by regions $A$ and $C$. In the absence of structural change, the mapped insert length is random with distribution $F$, which is chosen in the fragmentation step.  Fragments that span the deletion point, i.e. those that start in $A$ and end in $C$ in the target, produce read pairs that map further apart in the reference than expected under $F$.  Now consider Figure \ref{fig:schematic} (b), where an insertion $B$, spanned by $A$ and $C$, starts at position $t$ in the reference genome.  Reads that overlap with $B$ would fail to map, if $B$ were a foreign sequence with no homolog in the target genome, or it would map far from its mate, if $B$ were a ``domestic'' insertion from a distant location of the reference.  Read pairs where one read maps successfully and the other fails to map, maps in the same orientation, or maps too far from the first are called \emph{hanging pairs}.  Some alignment algorithms allow only the prefix or the suffix of a read to be mapped, often called ``soft-clipping.''  In this case, we also include read pairs where one read is soft-clipped in our definition of hanging pairs.   Deletions can also produce hanging pairs, if one read of a given pair straddles the boundary between $A$ and $C$ in the target genome.  Similarly, fragments that contain insertions produce read pairs that map closer to each other than expected under $F$.  In Section \ref{sec:model}, we  describe models and statistics that exploit these patterns to detect structural variants.

\subsection{Detection of Transcription Factor Binding Sites} \label{sec:chipseq}
Chromatin immuno-precipitation (ChIP) is a technique for isolating from a
DNA sample only those DNA fragments bound to a protein of interest.
Sequencing reads from the ends of the DNA fragments derived from ChIP, called
ChIP-Seq, then mapping these reads to a reference template allows us to detect
the binding locations of the protein in the genome of the sample. One expects
to see an increase in the density of mapped reads near the binding site.
Under the assumption that the binding site is short compared to $R$,
it is natural to assume that the ``shape'' of the peak centered on the site is roughly triangular.
Following Schwartzmann {\sl et al.} (2012), we consider
\begin{equation} \label{eq:chipseq}
Z_\tau = \int g_w(\tau - s) dN_s,
\end{equation}
where now it is convenient to assume that the jumps in $N$ are located at
the central nucleotides of the reads and $g$ is a symmetric kernel.
The function $g_w(s) = (1- |s|/w)^+/w$ is a plausible ``matched filter.''  An alternative kernel is a Gaussian probability density function with standard deviation $w$.  The scale parameter $w$ indicating
the width of the signal may  be known or unknown. It is easy to see that the log likelihood ratio for testing the intensity function $\rho(t)$ against the alternative of a peak at $\tau$
of the form $\exp(\beta g_w(\tau - s))$ equals
\begin{equation} \label{chipseq}
\ell(\tau, w,\beta) = \beta Z_\tau - \int [\exp(\beta
g_w(\tau - t) - 1]\rho(s) ds.
\end{equation}
Since the location $\tau$ is unknown, one might consider one of several statistics maximized over candidate values of $\tau$.  The simplest would be the score statistic,
$\partial \ell/\partial \beta |_{\beta = 0} = Z_\tau - \int g_w(\tau - s)
\rho(s) ds.$  An alternative would be $\max_\beta  \ell(\tau, w, \beta)$, where the maximum is over (in addition to $\tau$)  some appropriate range of values of $\beta > 0$
and perhaps also $w$.  Below we shall see other examples involving
convolution of a smooth function with a Poisson process.

\subsection{Modeling overdispersion}

It has often been found that the coverage process $N_t$ is overdispersed.  This can be handled by
using a negative binomial process, or equivalently a gamma mixture
of Poisson processes.  To see how this effects (\ref{deletion}),
consider a Poisson distribution with mean $\xi$, where
$\xi$ has the probability density function $[\Gamma(q)]^{-1} \alpha^q
\xi^{q-1} \exp(-\alpha \xi)$.  To maintain a baseline intensity of
$\rho(t)$ for
the observed process, we put $q = \alpha \Omega(t)$, where
$\Omega(t) = \int_0^t \rho(s) ds.$
The log likelihood
ratio for (\ref{deletion}) is
\begin{equation} \label{negbinom}
\beta  [N_{t_2} - N_{t_1-R}] + \alpha [\Omega(t_2) - \Omega(t_1 - R)]
\log[( 1 - \alpha^{-1} (e^\beta-1)],
\end{equation}
Note that if $\alpha \rightarrow \infty$, (\ref{negbinom})
converges to (\ref{deletion}).

\section{Models and Scan Statistics} \label{sec:model}
\subsection{A simple mixture model for structural variants} \label{sec:simple}
Consider first a simplified model for the detection of insertions
and deletions using the mapped insert lengths in paired-end sequencing.  We consider for now only those pairs where both reads are unambiguously mapped in opposite orientation.  For read pair $i$,
let $x_i^+$ and $x_i^-$ be the mapped positions of the plus and minus
strand reads, respectively.  For a reference template of length $T$, $(x_i^+,x_i^-) \in \{1,\dots, T-R+1\}^2$.  The mapped insert length, which we denote by $y_i$ for read pair $i$, is defined by $y_i \equiv x_i^- - x_i^+$.

If there are no structural variants, $y_i$ has distribution $F_0(dy)$ with
mean $\delta$ and standard deviation $\sigma$.  As described in Section \ref{sec:sv}, deletions cause an increase in mapped insert length, and small insertions cause a decrease.  We introduce a parameter $w$, where the sign of $w$ is positive for deletions and negative for insertions, and $|w|$ is the number of bases in the deleted or inserted segment.  Also, let $r$ be the proportion of genomes
in the sample carrying the variant.  Both $w$ and $r$ are usually unknown, although it will be convenient to study statistics defined by particular
values of these parameters.  Then, for read pairs straddling the deletion
point or inserted segment in the target, their mapped insert lengths
have the mixture distribution $F_1(dy) = (1-r)F_0(dy) + r F_0(dy-w).$

To detect insertions and deletions, we consider as a toy model the
two-dimensional Poisson random field
$$ N(dt,dy) = \sum_{i=1}^n I(x_i^+\in dt, y_i \in dy).$$
For simplicity, we assume for $N$ the null intensity function $\lambda(dt,dy) = \rho(t)dtF_0(dy)$, where $\rho(t)$ is the rate with which plus strand reads map to genome position $t$.  Alternatively, we can think of this process as
a marked (or compound) Poisson process with rate $\rho(t)$ and marks that follow $F_0$ or $F_1$.

An insertion or deletion starting at $s$ causes those read pairs with plus strand read mapping to a window before $s$ to have insert length following $F_1$.
This logic prompts the construction of an alternative intensity function \begin{equation}\label{lambda1simple}\lambda_1(dt,dy) = \left\{
                                      \begin{array}{ll}
                                        \rho(t)dtF_1(dy), & t \in [s-\delta,s); \\
                                        \rho(t)dtF_0(dy), & \hbox{otherwise.}
                                      \end{array}
                                    \right.\end{equation}
The log likelihood ratio of $\lambda_1$ versus $\lambda_0$ is
\begin{equation} \label{loglik1}
\ell = \int \log (\lambda_1(dt,dy)/\lambda_0(dt,dy)) N(dt,dy) - \int [\lambda_1(dt,dy) - \lambda_0(dt,dy)],
\end{equation}
The log likelihood is indexed by the parameters $s$, $w$, and $r$.  A scan of the
genome for large values of the log likelihood, varying $s$ and possibly
also $r$ and $w$, can be used to detect insertions and/or deletions.

Compared to models that we will introduce in Section \ref{sec:bettermodel}, the scan statistic
suggested here has a simple, general structure due to the assumption that
the rate function for the two dimensional process is a product of
one dimensional rate functions.  This leads to relatively simple
theoretical properties that may be of general interest for problems
involving mixtures in compound Poisson processes.
In regard to DNA structural variants, the
formulation ignores some important features of DNA sequencing data
by emphasizing information in the insert length and ignoring information
from the
coverage process and the hanging pairs, as described
in Section \ref{sec:sv}.  In particular the alternative rate function
(\ref{lambda1simple}) makes the simplifying
assumption that all read pairs with plus
strand read mapping within $[s-\delta,s)$ are equally informative about
the existence of a deletion/insertion at $s$.  In
Section \ref{sec:bettermodel}, we describe a more
realistic model for the structural variant detection problem.

\subsection{General framework and notation} \label{sec:general}
Before introducing more explicit models, we shall describe an abstract
framework for scans of Poisson-type data.  We assume that the observed data
are a counting process $\{N(dz): ~z \in \Omega\}$, that has a null
intensity function $\lambda_0(z)$ on the domain $\Omega$.  For example,
in the single-read sequencing set-up of Sections \ref{sec:cnv}
and \ref{sec:chipseq}, $N(z)$ is the coverage process, with $z$ being a
one-dimensional index for genome location.  In the mixture model proposed
in Section \ref{sec:simple}, $z=(t,y)$, $\Omega = [0,T]\times \Re$,
and $N(z)$ counts the number of read pairs with plus strand read mapping to a given location and mapped insert length within a given range.  The
signal of interest in all cases is a local change in intensity, which
is represented with an alternative intensity function $\lambda_1(z)$ that relies on one or more parameter(s), collectively denoted by $\tau$.  For example,
in Section \ref{sec:simple} $\tau$ can be the single parameter for genome
location, $s$, but can also be the vector $(s,r,w)$ which quantifies also the proportion and length of the signal.  For reasons that will
become apparent in Section \ref{sec:tailapprox}, we introduce the representation
\begin{equation} \label{lambda1}
\lambda_1(dz) = e^{\beta k_{\tau}(z)}\lambda_0(dz),
\end{equation}
where we call $k_{\tau}(z)$ the kernel function.  The parameter $\beta$ plays a
technical role in false positive rate calculations.  The alternative of
interest for the models in Sections \ref{sec:simple} and \ref{sec:bettermodel} is $\beta=1$.  Expressed in this way, (\ref{loglik1}) can
be written somewhat more abstractly in the form
\begin{equation} \label{loglik}
\ell_\tau = \beta \int k_\tau(z) N(dz) - \psi_{\tau}(\beta),
\end{equation}
where
\begin{equation} \label{psi}
\psi_{\tau}(\beta)=\int \{\exp[\beta k_\tau(z)]-1\} \lambda_0(dz)\;.
\end{equation}
It is easy to see that the scan statistics (\ref{deletion},\ref{chipseq})
for detecting copy number variants and peaks in ChIP-Seq data can be written in this form.  The representation (\ref{loglik}) allows for much simpler
notation in moment calculation and tail approximations in Sections
\ref{sec:moments} and \ref{sec:tailapprox}.

To complete the specification of the alternative intensity
(\ref{lambda1simple}), we assume that
$F_0$ is given by the normal density with respect to
$y$, so $\lambda_0(dt,dy) = \rho(t) dt \phi(y)dy$.  The normal distribution
has been a good fit to the center of the mapped insert length distributions in the data that we have examined; see detailed example in Section \ref{sec:data}. For the mixture model, we put
$\tau = (s,r,w)$ and set
\begin{equation}\label{mixturekernel}k_\tau(t,y) =
1\{s- \delta \leq t \leq s\} \log(1+r[\phi(w - y)/\phi(y) -1]). \end{equation}
%A different possibility is $k_\tau(t,y) = 1(s\leq t \leq s+L) \exp\{-(y-w)^2/(2r^2)\}$.
To emphasize that $k_{\tau}(t,y)$ is separable and to simplify certain expressions given
below, we often write (in obvious notation)
\begin{equation} \label{sep}
k_\tau(t,y) =
1\{s - \delta  \leq t \leq s\} g(y;w, r).
\end{equation}
In both of these cases the rate of the baseline Poisson process under the
formal alternative is identical to the rate of the null model outside the
interval $[s-\delta, s]$.

In general we will consider, as raw material for scan
statistics, the random fields (\ref{loglik}), indexed by
the unknown  parameters $\tau$.   Note that the random
field is in general not differentiable in  the parameter $s$
for location, but is typically differentiable in the
other parameters that determine the alternative distribution.
%and the efficient score
%with respect to $\beta$ evaluated at $\beta = 0$, which equals
%\begin{equation} \label{score}
%Z_\tau = \int k_\tau(u) [N(du) - \lambda_0(du).
%\end{equation}

%Note that the second integral in (\ref{loglik}) is the cumulant generating
%function of $Z_\tau$, evaluated at $\beta$.

\subsection{More realistic models for structural variants} \label{sec:bettermodel}
The mixture model suggested above neglects a number of features of the problem of detecting structural variants by paired end reads.  Here we propose alternative models, which differ slightly between insertions and deletions.

Let $n$ be the number of read pairs where at least one read within the pair is successfully mapped to the template.    Note that in Section \ref{sec:simple}, only those pairs where \emph{both} reads are successfully mapped were considered.  As before, let $x^+_i$ and $x^-_i$ be the left most base positions of the plus and minus strand reads, respectively, for pair $i$.  Successfully mapped reads have positions in $\{1,\dots,T-R+1\}$.  In all DNA sequencing experiments, some reads will fail to map to the reference template, in which case we assign its position the value $\infty$.  Reads may fail to map due to sequencing or mapping error, or due to its inclusion of a segment of DNA that does not have a match in the reference.  Read pairs where the plus (minus) strand failed to map are called \emph{hanging plus (minus) strand pairs}.  Let $p$ be the probability of a hanging pair due to experimental error.  A conservative estimate of $p$ can be obtained by $n^{-1}\sum_i [I(x_i^+=\infty)+I(x_i^-=\infty)].$

In Section \ref{sec:sv} we defined hanging pairs more broadly, so that it includes also pairs that are mapped too far apart, in reverse orientation, or pairs that include soft-clipped reads.  The models and statistics we introduce below easily adapt to the broader definition, but the notation will be much simpler under the narrow definition.  The important thing is that, given the definition for a hanging pair, $p$ must be empirically estimated by the proportion of such hanging pairs among all read pairs with at least one read mapped.

%A read pair is considered ``properly mapped'' if both reads are successfully mapped to the template, and if the mapped insert length is within some pre-determined probable range.  Departing from our previous model, we assume that the density function $f(y)$ for insert lengths has support $[\delta-\Delta, \delta+\Delta]$, reflecting the belief that fragments beyond this size range are improbable.  Improper pairs are caused either by errors in sequencing or mapping, or by an insertion or deletion straddled by the pair.  We let $p$ be the probability of an improper pair caused by error.  When there is an error, we assume that the insert length is chosen uniformly from the possible range outside of the support of $f(y)$.

Let $\kappa(t)$ be the rate with which reads (either plus or minus strand) map to position $t$.  Although reads map to integer positions, as before, for mathematical convenience we embed the mapping positions in to the continuous interval $[0, T]$ and let
\begin{equation}
    N(du,dv)= \sum_{i=1}^n I(x_i^+ \in du, ~x_i^-\in dv), \quad u,v \in [0,T].
\end{equation}
Then, in the notation of Section \ref{sec:general}, $N$ is an inhomogeneous Poisson Process with $z=(u,v)$, $\Omega = ([0,T] \cup \infty)^2$, and intensity function
\begin{equation}
\lambda_0(u,v)= \left\{
                  \begin{array}{ll}
                    (1-p)\kappa(u)\kappa(v)f(v-u), & u,v \in [0,T];\\
                    \frac{1}{2}p\kappa(u)\int_{u}^T \kappa(x) f(x-u)dx, & u \in [0,T],~v=\infty;\\
                    \frac{1}{2}p\kappa(v) \int_0^v\kappa(x) f(v-x)dx, & u=\infty,~v \in[0,T].
                  \end{array}
                \right.
\end{equation}
The integrals in the second and third lines account for the possible
different insert lengths, which are unobserved because of
the hanging read.  We assume that hanging pairs have probability half for each of plus strand and minus strand read hanging.
Note that the marginal intensity for a read to map to $t$ is
$\kappa(t)$.  If we assume constant $\kappa$, then $\lambda_0(u,v)$ simplifies to $(1-p)\kappa^2f(v-u)$ for properly mapped read pairs, and $p\kappa^2/2$ for plus and minus strand hanging pairs.

Now consider testing the alternative hypothesis that a proportion $r$ of
the genomes in the sample contain a deletion of width $w$ beginning at
reference location $s$.  In reference to the window $[s,s+w)$,
the sample space $\Omega$ can be partitioned into the following non-overlapping sets:
\begin{eqnarray*}
S^C=S^{C}_{s,w} &=& \{(u,v):~s-R<u < s+w ~\hbox{or}~ s-R<v < s+w \};\\
S^B=S^{B}_{s,w} &=& \{(u,v):~ u \leq s-R ~\hbox{and}~  v > s+w\};\\
S^+=S^+_{s,w} &=& \{(u,v):~ u =\infty  ~\hbox{and}~  v> s+w\};\\
S^-=S^-_{s,w} &=& \{(u,v):~ u\leq s-R ~\hbox{and}~  v =\infty\};\\
S^0=S^0_{s,w} &=& \Omega\backslash (S^{C}_{s,w} \cup S^{B}_{s,w} \cup S^{+}_{s,w} \cup S^{-}_{s,w}).
\end{eqnarray*}
$S^{C}_{s,w}$ is the set of pairs where at least one read
intersects the window $[s,s+w)$; $S^{B}_{s,w}$ is the set of pairs that \emph{b}racket the window; $S^{+}_{s,w}$ is the set of hanging plus strand pairs where the minus strand read maps to the right of the window; $S^{-}_{s,w}$ is the set of hanging minus strand pairs where the plus strand read maps to the left of the window; $S^{0}_{s,w}$ contains all of the remaining pairs, which are uninformative about whether there is a deletion of $[s,s+w)$.  Under broader definitions of hanging pairs, the definition for these sets can be easily adjusted so that they remain a partition of $\Omega$.  To simplify notation we will sometimes suppress the suffix $s,w$.

Let $\lambda_1(u,v)$ be the rate function under the alternative of a
deletion with parameters $\tau=(s,w,r)$.  To specify $\lambda_1$, we
consider the probability under the alternative of read pairs belonging to
each of the above sets separately.  The deletion should not affect the rate
with which pairs map to $S^0$.  Pairs in $S^C$ can only come from
the non-carrier genomes, with probability $1-r$, and thus \begin{equation} \label{del1}\lambda_1(u,v) = \lambda_0(u,v)[1-r],\quad (u,v) \in S^C.\end{equation}   A pair in $S^B$ can be generated in two ways:  It can be from a non-carrier
chromosome, with rate $(1-r)\lambda_0(u,v)$, or it can be from a fragment
containing the deletion from the carrier chromosome, with rate
$r(1-p)\kappa(u)\kappa(v)f(v-u-w)$.  Thus,
\begin{equation}\label{del2}
\lambda_1(u,v) = \lambda_0(u,v)[1-r+rf(v-u-w)/f(v-u)], \quad (u,v) \in S^B.
\end{equation}
Now consider the hanging minus strand pairs.  A pair mapping to $(u,v)\in S^-$ can be from a non-carrier chromosome, with rate $(1-r)\lambda_0(u,v)$, or it can be from a carrier chromosome.  In the latter case, there are two explanations for the minus strand read failing to map: It can be due to sequencing error, or it can be due to the read overlapping the deletion point.  Thus, for $(u,v) \in S^-$,
\begin{eqnarray}
\lambda_1(u,v) &=& (1-r)\lambda_0(u,v) + r[\lambda_0(u,v)+(1-p)\kappa(u)\int_{s-R}^sf(t-u)\kappa(t)dt] \nonumber \\
&=& \lambda_0(u,v)\left[1+\frac{2r(1-p)}{p\int_{u}^T \kappa(x) f(x-u)dx}\int_{s-R}^{s}f(t-u)\kappa(t)dt\right]. \label{del3}
\end{eqnarray}
With similar reasoning, we have for $(u,v)\in S^+$
\begin{eqnarray}
&& \label{del4}\\\lambda_1(u,v)
&=& \lambda_0(u,v)\left[1+\frac{2r(1-p)}{p\int_0^v\kappa(x) f(v-x)dx}\int_{s+w-R}^{s+w}f(v-t)\kappa(t)dt\right].  \nonumber
\end{eqnarray}
It is easy to see that the alternative rate function can be written in the form of (\ref{lambda1}) with $\beta=1$, $k_{\tau}=0$ for $(u,v)\in S^0$, and $k_{\tau}$ equal to the log of the term in square brackets in (\ref{del1}-\ref{del4}) for $(u,v)$ belonging to, respectively, $S^C$, $S^B$, $S^+$ and $S^-$.  The log-likelihood scan statistic thus evaluates to
\begin{equation}\label{likdel}
\ell_{\tau} = \beta[Z^C_{\tau} + Z^B_{\tau} +Z^+_{\tau}+ Z^-_{\tau}] - \psi_{\tau}(\beta),
\end{equation}
where $Z^C_{\tau}$, $Z^B_{\tau}$, $Z^+_{\tau}$ and $Z^-_{\tau}$ are the sum of of the kernel $k_{\tau}$ over the sets $S^C$, $S^B$, $S^+$, and $S^-$, respectively.  That is,
\begin{eqnarray*}
Z^C_{\tau} &=& \sum_{i:(x^+_i,x^-_i)\in S^C} \log(1-r); \\
Z^B_{\tau} &=& \sum_{i:(x^+_i,x^-_i)\in S^B} \log[1-r+rf(x^-_i-x^+_i+w)/f(x^-_i-x^+_i)];\\ Z^+_{\tau} &=& \sum_{i:(x^+_i,x^-_i)\in S^+} \log\left[1+\frac{2r(1-p)\int_{s+w-R}^{s+w}f(x^-_i-t)\kappa(t)dt}{p\int_0^{x^-_i}\kappa(x) f(x^-_i-x)dx}\right];\\
Z^-_{\tau} &=& \sum_{i:(x^+_i,x^-_i)\in S^-} \log\left[1+\frac{2r(1-p)\int_{s-R}^{s}f(t-x^+_i)\kappa(t)dt}{p\int_{x^+_i}^T \kappa(x) f(x-x^+_i)dx}\right].
\end{eqnarray*}
We call $Z^C_{\tau}$, $Z^B_{\tau}$, $Z^+_{\tau}$, and $Z^-_{\tau}$ signature specific scores, or simply, scores, since they summarize the evidence for a deletion from, respectively, the coverage process, the bracketing pairs, the hanging plus strand pairs, and the hanging minus strand pairs.  If $\kappa$ were assumed constant, the scores for the hanging pairs simplify significantly to
\begin{eqnarray*}
Z^+_{\tau} &=& \sum_{i:(x^+_i,x^-_i)\in S^+} \log\{1+\frac{2r(1-p)}{p}[F(x_i^--{s+w-R})-F(x_i^--{s+w})]\},\\
Z^-_{\tau} &=& \sum_{i:(x^+_i,x^-_i)\in S^-} \log\{1+\frac{2r(1-p)}{p}[F(s-x_i^+)-F(s-R-x_i^+)]\}.
\end{eqnarray*}
From these simplified versions, we see that the hanging pairs scores are weighted counts of the hanging pair of the given type in the region before the start of the deletion (for $Z^-$) or after the end of the deletion (for $Z^+$), where the weights depend on the insert length distribution $F$.

The above reasoning can be easily modified to handle insertions.  For testing the alternative of an insertion of width $w$ between template positions $s$ and $s+1$ in a proportion $r$ of the chromosomes, we redefine the sets
\begin{eqnarray*}
S^{C}_{s,w} &=& \{(u,v):~s-R<u \leq s ~\hbox{or}~ s-R<v \leq s \};\\
S^{B}_{s,w} &=& \{(u,v):~ u \leq s-R ~\hbox{and}~  v > s\};\\
S^+_{s,w} &=& \{(u,v):~ u =\infty  ~\hbox{and}~  v>s\};\\
S^-_{s,w} &=& \{(u,v):~ u\leq s-R ~\hbox{and}~  v =\infty\};\\
\end{eqnarray*}
Then, $\lambda_1(u,v)$ remains the same as (\ref{del1}) for $S^C$, and the same as (\ref{del2}) with $-w$ replaced by $+w$ for $S^B$.  For the hanging minus strand pairs,
\begin{equation}
\lambda_1(u,v) = \lambda_0(u,v)[1+\frac{2r(1-p)}{p\int_{u}^T \kappa(x) f(x-u)dx}\int_{s-R}^{s+w}f(t-u)\kappa(t)dt],
\label{ins3}\end{equation}
and for the hanging plus strand pairs,
\begin{equation}
\lambda_1(u,v) = \lambda_0(u,v)[1+\frac{2r(1-p)}{p\int_0^v\kappa(x) f(v-x)dx}\int_{s-w-R}^sf(v-t)\kappa(t)dt].\label{ins4}
\end{equation}

\begin{remark} \label{rem:alignpeaks}
There is an important
difference between insertions and deletions for the hanging
read statistic.  For insertions both $Z^+$ and $Z^-$ should give a peak at the point of the insertion
in the reference genome, hence can be combined by addition.
For deletions of the interval $(s,s+w)$, $Z^-$ should give a peak at $s$, while $Z^+$ should give a peak at $s+w$.  These
two statistics will reinforce each other if $w$ is small
enough for the two peaks to overlap.  Since $w$ is unknown,
alignment can be accomplished by maximizing the sum of the
two statistics over a range of $w$ values, which must be paid
for by a larger significance threshold.   As we shall see,
the hanging read statistics
are most useful for detecting short indels, where the bracketing
pairs statistics have little power.   The ideal range depends on
the true value of $w$ and on
other unknown parameters.  For simplicity in what follows we
carry out the maximization over [0, 150].  Under some conditions
eiher a shorter or a longer range might be better.
\end{remark}

This model, although much more precise than the toy model in Section \ref{sec:simple}, still cuts some corners.  One tricky issue is that the rate function $\kappa(t)$, which
reflects the ease of fragmentation and mapping, is not known
for $t$ near deletion points or for $t$ within inserted segments.
In (\ref{del3}, \ref{del4}, \ref{ins3}, and \ref{ins4}) we ignored this
issue and used an incorrect value of $\kappa$.  In practice, it suffices to
replace $\kappa(t)$ with a genome-wide average in the summands
of these formulas.

It is not a priori clear whether one should try to combine the scores $S^C$, $S^B$, $S^+$ and $S^-$ into a single statistic, as in
$\ell_{\tau}$, or treat them separately, e.g.,
by a scan with only $Z^B$ to target relatively long intervals
and $Z^++Z^-$ to target short intervals, then use
a Bonferroni bound to correct for using two different statistics.  In
Section \ref{sec:power} we will explore the sensitivity
of the various types of scans.

\begin{remark}
Like the simplified model of $S^B$ proposed in Section \ref{sec:simple},
it is also possible to develop a simplified model for the
``hanging read'' statistics, $S^+$ and $S^-$.  If we assume that
there is no variability in the insert lengths, i.e., $\sigma = 0$,
then for a mapped positive strand read beginning in the interval
$[s-\delta, s-\delta + R]$ the corresponding negative
strand will not map (a)  whenever there is a deletion
beginning at $s$  or (b) with probability
$p$, even if there is no deletion.  Hence a simple
detection statistic would be obtained by
counting the number of reads beginning in each interval of length $R$,
the other end of which does not map,
and claiming a detection of an deletion at $s$  whenever the sum of
positive strand reads that begin
$[s-\delta, s-\delta + R]$ and negative strand reads that begin in
$[s+\delta - R, s+\delta]$ is too large to be a determined by chance.
An appropriate modification would serve to detect insertions.
Some numerical experimentation
suggests that this simplified test is less powerful than the
more detailed likelihood based procedure described above.
A numerical example is contained in Section 8.
\end{remark}

\section{Moments} \label{sec:moments}

In this and the following sections we develop a method for computing
approximate p-values for scans of the form (\ref{loglik}).  Our approach relies on a measure transformation technique that shifts the distribution towards the desired alternative within the scan window.  See Siegmund, Yakir, and Zhang (2012), Yakir (2013), and references cited there for details of this method and
its applications to several different problems.

We begin by deriving the moments of the likelihood ratio statistic under measure transformations, which will be useful for p-value approximations and power calculations.  Consider the expectation and variance of (\ref{loglik}).
%Under the null
%distribution the rate of the poisson field is $\lambda$. Hence:
%\[
%\Expec[\ell_t] = \int k_t(u) \Expec[N(du)] - \int [e^{k_t(u)}-1]
%\lambda(du) = \int [ k_t(u) - e^{k_t(u)} + 1] \lambda(du)\;.
%\]
%The variance under the null distribution is:
%\[
%\Var[\ell_t] = \int [k_t(u)]^2 \Expec[N(du)] = \int
%[k_t(u)]^2\lambda(du)\;.
%\]
Let $\Prob$ be the measure where $N(dz)$ has null intensity $\lambda_0(dz)$, and define
$$ d\Prob_{\tau} = \exp(\ell_{\tau})d\Prob = \exp\left[\beta \int k_\tau(z) N(dz) - \psi_{\tau}(\beta) \right]dP.$$
It's easy to show that under $\Prob_{\tau}$, $N(dz)$ is still a Poisson random field but with intensity function (\ref{lambda1}).  Let $\Expec_{\tau}$ and $\Var_{\tau}$ be expectation and variance, respectively, under $\Prob_{\tau}$.  Then, the first two moments of $\ell_\tau$ under the alternative measure are
\begin{eqnarray*}
\Expec_\tau[\ell_\tau] &=& \int \beta  k_\tau(z) \Expec_\tau[N(dz)] - \int [e^{\beta k_\tau(z)}-1]
\lambda_0(dz)\\ &=& \int [ \beta k_\tau(z)e^{\beta k_\tau(z)} - e^{\beta k_\tau(z)} + 1]
\lambda_0(dz)\;,
\end{eqnarray*}
and
\[
\Var_\tau[\ell_\tau] = \int [\beta k_\tau(z)]^2 \Expec_\tau[N(dz)] = \int
[\beta k_\tau(z)]^2e^{\beta k_t(z)}\lambda_0(dz)\;.
\]
For the mixture model in Section \ref{sec:simple}, when $k$ is written in the form (\ref{sep}), these simplify to
\[
\Expec_\tau[\ell_\tau] = \int_{t_1}^{t_2} \rho(t)dt \int [\beta g(y,w,r) \exp[\beta g(y,w,r)]
- \exp[\beta g(y,w,r)] + 1] dF_0(y)
\]
and a similar expression for the variance.

%The gradient of the expectation with respect to $\tau$ is
%\begin{equation} \label{grad}
%\int \beta^2 \dot k_\tau(u) k_\tau(u)e^{\beta k_\tau(u)}\lambda_0(du).
%\end{equation}
%which specifies for the normal kernel to be:
%\begin{align*}
%\frac{\partial}{\partial \mu} \Expec_t[\ell_t] =
%\frac{\beta^2}{\sigma^2} \rho (\tau_2-\tau_1)\int
%(y-\mu)\tilde g_t(y) dy &\propto \frac{\beta}{\sigma^2}(\tilde \mu_t - \mu)\;,\\
%\frac{\partial}{\partial \sigma^2} \Expec_t[\ell_t] =
%\frac{\beta^2}{2\sigma^4} \rho (\tau_2-\tau_1)\int
%(y-\mu)^2\tilde g_t(y) dy &\propto \frac{\beta}{2\sigma^4}\{\tilde \sigma_t^2+(\tilde \mu_t - \mu)^2\}\;,\\
%\frac{\partial}{\partial \beta} \Expec_t[\ell_t] = \beta \rho
%(\tau_2-\tau_1)\int \tilde g_t(y) dy &\propto 1\;,
%\end{align*}
%where $\tilde \mu_t$ and $\tilde \sigma^2_t$ are the expectation and
%variance of the distribution that has a density proportion to:
%\[
%\tilde g_t(y) = \frac{1}{\sqrt{2\pi}}\exp\Big\{-(y-\mu)^2/\sigma^2 -

\subsection{The expectation and covariance structure of the local field} \label{sec:localmoments}
We call $l_{\sigma}-l_{\tau}$, for $\sigma$ close to $\tau$, the local field of $\tau$.  For p-value approximations we will also need the moments of the local field under $\Prob_{\tau}$.  The expectation is
$$
\Expec_\tau(\ell_\sigma - \ell_\tau) = \int \big[ \beta (k_\sigma(u) - k_\tau(u)) e^{\beta k_\tau(u)}
- e^{\beta (k_\sigma(u))} + e^{\beta (k_\tau(u)}\big] \lambda_0(du)\;.
$$
Now consider the mixture model where $k$ is in the form (\ref{sep}), where $\tau =(t_1,t_2,r,w)$ and $\sigma = (t_1,t_2,q,\eta)$.  Let $D_{\sigma}$ denote differentiation with respect to $\sigma$, and let $\dot{k}$ denote partial derivatives of $k$ with respect to
$(q,\eta)$. Then
\[
%\bigtriangledown \Expec_\tau(\ell_\sigma - \ell_\tau) = \int
D_{\sigma}\Expec_{\tau}(\ell_{\sigma} - \ell_{\tau}) = \beta\int
\big[\dot  k_{\sigma}(u) \big(e^{\beta k_{\tau}(u)} - e^{\beta k_{\sigma}(u)}\big) \big]
\lambda(du)\;,
\]
which vanishes when $s=t$. The Hessian is
\[
%{\bigtriangledown}^2 \Expec_\tau(\ell_\sigma - \ell_\tau) = \int
D^2_{\sigma}\Expec_{\tau}(\ell_{\sigma} - \ell_{\tau}) = \int
 \big[\beta \ddot  k_{\sigma}(u) \big(e^{\beta k_{\tau}(u)} - e^{\beta k_{\sigma}(u)}\big) \big] - \beta^2[\dot  k_{\sigma}(u)] [\dot  k_{\sigma}(u)]' e^{\beta k_{\sigma}(u)} \big]
\lambda(du)\;.
\]
Evaluation when $\sigma=\tau$ produces:
\[D^2_{\sigma}\Expec_{\tau}(\ell_{\sigma} - \ell_{\tau})  \Big
|_{\sigma=\tau} = - \beta^2 \int [\dot  k_{\tau}(u)] [\dot  k_{\tau}(u)]' e^{\beta k_{\tau}(u)}
\lambda(du) = - \Sigma_{\tau}\;.
\]

Considering the random vector $\dot \ell_{\tau} = \beta \int \dot k_{\tau}(u)
N(du)$. The covariance matrix satisfies $\Var_{\tau}(\dot
\ell_{\tau}) = \Sigma_{\tau}$. Clearly, this is also the variance-covariance
matrix of the random gradient of the centered process $(\ell_{\sigma} -
\ell_{\tau}) - \Expec_{\tau}(\ell_{\sigma} - \ell_{\tau})$.

If the values of $w$ and $r$ and $\beta$ are held fixed and
the values of $\tau_1$ and $\tau_2$ are allowed to vary, we
should consider the two parameter sets \begin{equation} \label{mmlocinc} \tau =
(t_1,t_2,w,r,\beta), \quad \sigma =
(t_1+\epsilon_1,t_2+\epsilon_2, w,r,\beta). \end{equation} The
expectation of the difference is
\[
\Expec_{\tau}(\ell_{\sigma} - \ell_{\tau}) = \frac{\epsilon_2^+ +
\epsilon_1^-}{t_2-t_1}\Expec[\ell_{\tau}] - \frac{\epsilon_2^- +
\epsilon_1^+}{t_2-t_1}\Expec_{\tau}[\ell_{\tau}]
\]
and the variance is
\[
\Var_{\tau}(\ell_{\sigma} - \ell_{\tau}) = \frac{\epsilon_2^+ +
\epsilon_1^-}{t_2-t_1}\Var[\ell_{\tau}] + \frac{\epsilon_2^- +
\epsilon_1^+}{t_2-t_1}\Var_{\tau}[\ell_{\tau}]\;.
\]
This local process has independent increments.

In the special case where $\epsilon_1 = \epsilon_2 = \epsilon$ the
expectation reduces to
\begin{equation}\label{xi}
\Expec_{\tau}(\ell_{\sigma} - \ell_{\tau}) =
\frac{\Expec[\ell_{\tau}]-\Expec_{\tau}[\ell_{\tau}]}{t_2-t_1}\cdot |\epsilon
| = -\xi_{\tau} \cdot |\epsilon |,
\end{equation}
and the variance reduces to
\begin{equation} \label{varsigma}
\Var_t(\ell_s - \ell_t) =
\frac{\Var[\ell_t]+\Var_t[\ell_t]}{t_2-t_1}\cdot |\epsilon | =
\varsigma^2_{\tau} \cdot |\epsilon |\;.
\end{equation}

\section{The probability of crossing a threshold} \label{sec:tailapprox}

Consider, for some threshold function $x_t$, $\Prob(\cup_{t\in T}
\{\ell_t \geq x_t\})$, the probability that the likelihood ratio statistic (or some other suitable scanning statistic) crosses the threshold at some point in $T$. We assume that the threshold is
high enough so the probability converges to zero but, on the other
hand, it is not too high to allow for the application of local and
nonlocal central limit theorems where appropriate. We propose to use
the following steps in order to produce an analytical approximation
for this probability:

\begin{enumerate}
\item Identify the parameter value(s) that maximize the marginal
probability:
\[
\tau = \arg\max_{t\in T}  \Prob(\ell_t \geq x_t)\;.
\]
\item Restrict $T$ to the collection of parameter values for which
the marginal probabilities are in the same order of magnitude as the
maximal marginal probability.

\item Apply the measure-transformation technique described in Siegmund, Yakir and Zhang (2011).
\end{enumerate}
The measure transformation technique relies on a rewriting of the probability of interest,
\begin{align} \label{mt}
\Prob(\cup_{t\in T} \{\ell_t \geq x_t\}) = \sum_{t\in
T}e^{-x_t}
\Expec_t\Big[\frac{M_t}{S_t}e^{- [\ell_t-x_t+m_t]}; \ell_t - x_t + m_t \geq 0 \Big]
\end{align}
with $M_t$ and $S_t$ being the maximization and summation,
respectively and with respect to $s$, of $\exp\{\ell_s-\ell_t +x_t -
x_s\}$ and $m_t = \log M_t$. Note that $M_t$ and $S_t$ rely only on the local field of $t$, whereas the rest of the quantity within $\Expec_t[\dots]$ above rely on the ``global'' field $l_t$.  The localization theorem (Theorem 3.1 of Siegmund, Yakir and Zhang (2011)) states that, under certain conditions, the local and global components are asymptotically independent, which gives
\[
\lim_{\rho \rightarrow
\infty}\rho^{\frac{1}{2}}\Expec_t\big[(M_t/S_t) e^{-(\ell_t -x_t +
m_t)}; \ell_t -x_t + m_t \geq 0\big] =
\sigma_t^{-1}\phi(\mu_t/\sigma_t)\Expec_t\big[\mathcal{M}_t/\mathcal{S}_t]\;,
\]
where \begin{equation}\label{mut} \mu_t = \lim_{\rho \rightarrow
\infty}[\Expec_t(\ell_t)-x_t]/\rho^{\frac{1}{2}}, \quad \sigma_t^2 =
\lim_{\rho \rightarrow \infty}\Var_t(\ell_t)/\rho.\end{equation}

To evaluate $\Expec\big[\mathcal{M}/\mathcal{S}]$ it should
be noted that in order to obtain stochastic convergence of the local
field one should differentiate between the smooth component and the
brownian motion-type component in $l_t$. For the former the appropriate rate parameter is
$\rho^{-\frac{1}{2}}$ and for the latter the it is
$\rho^{-1}$.  For example, considering the case where $\epsilon_1 =
\epsilon_2 = \epsilon$ and $u$ standing for local increments in $w$ and/or $r$ in (\ref{mmlocinc}), we take $s = t +
(\rho^{-1}k,\rho^{-1}k,\rho^{-\frac{1}{2}} u)$.
To determine the mean of the local field, we take second order Taylor expansion of $\mu_t$ with respect to the smooth parameter(s) $u$ while holding the Brownian motion-type parameters for genome location fixed, and then add that to the drift  (\ref{xi}) of the Brownian motion type component. In the notation of Section \ref{sec:localmoments}, we have
\[
\ell_s-\ell_t +x_t - x_s = \varsigma_tB(k)-k\xi_t + u'\dot {\ell_t}
- \frac{1}{2}u'\Sigma_t u + o_p(1)\;,
\]
where $B$ is a two-sided random walk and $\xi_t$, $\varsigma_t$ are defined in (\ref{xi}, \ref{varsigma}).  In particular, note that
elements in the second order expansion that involve products of Brownian
motion elements and smooth elements are negligible. It follows that:
\[
\Expec_t\big[\mathcal{M}_t/\mathcal{S}_t] =
\rho^{-\frac{5}{2}}\times \nu(2\xi_t/\varsigma_t^2) \times
(2\pi)^{\frac{3}{2}}|\Sigma_t|^{-\frac{1}{2}}\;.
\]
A variation on this method also applies when, as for example in equation (2),
the kernel $g$ is continuous.  The details are omitted here, but some are given
below in the examples.

\section{Analysis of Scan Statistic with Mixture Alternative}

The statistics we study here are derived from maximizing (\ref{loglik})
over a suitable range of genomic locations $t$, and optionally also a subset of the other parameters in $\tau$.

In the notation introduced earlier, let
\[
Z(t,w,r) = \int_y \int_{t- \delta}^{t}  g(y; w,r) N(ds, \; dy).
\]
The log likelihood is given by (cf.
(\ref{loglik}))
\[ \ell(t,w,r) = \beta Z(t,w,r) - \Omega_\delta(t) \; \psi(\beta;w,r),
\]
where $\Omega_\delta(t) =\int_{t- \delta}^{t} \rho(s) ds$ and $\psi(\beta; w,r)
= \int \{\exp[\beta g(y; w,r)]-1\}dF_0(y)$
is the cumulant generating function of $Z(t;w,r)/\Omega_\delta(t)$.

Straightforward calculations show that under the formal
alternative
\[ \Expec[Z(t;w,r)] / \Omega_\delta(t)=
\int \{g(y; w,r)\exp[\beta g(y; w,r)]\}dF_0(y)
 ~~~ \equiv \xi(\beta).
\]
When there is no signal the expectation is
$\xi(0)$.
Similarly the variance of $Z$ is $\Omega_\delta(t) $ times the quantity
$\sigma^2(\beta) = \int \{g^2(y; w,r)\exp[\beta g(y; w,r)]\}dF_0(y).$
The expected value of $\ell(t;w,r)$  can be expressed as
$\Omega_\delta(t) \; J(\beta, w,r)$, where
$ J(\beta, w,r) = [\beta D_\beta \psi - \psi]$,
with $D_\beta$ denoting differentiation with respect to $\beta.$  The
parameter $J$ is the
Kullback-Leibler information and plays an important role in our use of
exponential change of measures to control the false positive error rate.

We consider detection statistics of the form
\begin{equation} \label{truelr}
\max Z(t; w,r)
\end{equation}
and
\begin{equation} \label{impliedlr}
\max \ell(t;w,r).
\end{equation}
The maximum can extend over $(t,w)$ or over $(t,w,r)$ in some suitable range.
We assume that $t$ changes by discrete amounts $\Delta > 0$.
For the most part we take $\Delta = 1$, but for some theoretical
results it is useful to let $\Delta \rightarrow 0$.
One could consider arbitrary fixed values of $w$
and $r$, but power may be increased
by considering a range of values for $w$, say $[w_0, w_1]$.
We also consider
maximization over a range of values of $r$, but our power calculations show that this maximization does not give a clear boost in sensitivity.

\begin{remark}
The statistic (\ref{truelr}) is essentially
the scan statistic studied by Chan and Zhang (2007) and
applied to the problem of detecting origins of replication in
viral genomes.  Chan and Zhang, however, study specific
``scoring'' functions $g$ that are free of unspecified parameters.  The rate $\rho(t)$ is also held constant, and thus under a fixed window size (\ref{truelr}) is equivalent to (\ref{impliedlr}).
They do not consider a general maximum likelihood
analysis of alternatives to
the null model, and their calculations are equivalent to using
what we have called the formal alternative with the value
$\beta$ defined by  (\ref{implied1}) below.
\end{remark}

\subsection{Approximate p-values for homogeneous processes.} \label{sec:hommix}

For simplicity we assume that $\rho(t) = \rho $ for all $t$,
so $\Omega_\delta(t) = \rho \delta$ is independent of $t$. In this section
we suggest p-value approximations for the various
scan statistics discussed above.   Mathemtically precise
versions of these approximations are discussed later.  The approximations
suggested here are simpler to evaluate and appear to be
slightly conservative.

For fixed values of $w,r$, one can find approximately the tail probability
of (\ref{truelr}) or (\ref{impliedlr}) by application of the methods
of Chan and Zhang (2007) or Siegmund,
Yakir, and Zhang (2011).
For example, let
$\Expec[\ell(t; w,r)] = \rho \delta J(\beta; w,r)$
denote Kullback-Leibler information.  Assume $\beta$ is chosen, so that
\begin{equation} \label{implied1}
\rho \delta J(\beta; w,r) = x.
\end{equation}
Then for large $x$
and $\rho \delta$,
\[
\Prob_0 \{ \max_{1 \leq t \leq m}\ell(t;w,r) \geq x \}
\]
\begin{equation} \label{special}
\approx 1 - \exp\left\{- m e^{-x}\rho[\xi(\beta)- \xi(0)](2\pi \rho \delta)^{-1/2} \sigma(\beta)^{-1} \nu\left(\frac{2\rho^{1/2}[\xi(\beta)+ \xi(0)]}{[\sigma^2(\beta)+\sigma^2(0)]^{1/2}}\right)\right\},
\end{equation}
where $\nu$ is the function defined in Siegmund (1985) and
given approximately for
purposes of numerical evaluation in Siegmund
and Yakir (2007);  and
where we have for simplicity
assumed that $F_0$ is a non-lattice distribution.

The function $\nu(y)$ is always between 0 and 1 and approximately equals 1 for
small values of $y > 0.$  Although it appears that inclusion of
$\nu$ improves the quality of the approximation, in what follows we occasionally
take $\nu$ identically equal to one.  This simplifies some calculations, and
numerical experimentation indicates that it rarely affects the power by more
than a few per cent.

The corresponding approximations when we also maximize over $w$ or over
$w$ and $r$ are more complicated.
Consider, for example, the event
\begin{equation}  \label{lrinequality}
R = \{ \max_{t, w_0 \leq w \leq w_1} [\ell(t; w,r)-x_{w,r}]
\geq 0\}.
\end{equation}
Let $J(\beta_w,w,r) = \Expec[\ell(t;w,r)]/(\rho \delta)$, with
$\beta_w$ chosen
so that $\rho \delta J(\beta,w,r) = x_{w,r}$.
Then $\Prob_0(R) \approx$

\begin{equation} \label{generalapprox}
1- \exp\left\{-m\rho\int_{w_0}^{w_1}\frac{\exp(-x_{w,r})
[\xi(\beta_w)- \xi(0)]
\nu\{ \cdot \}
[\Sigma(w)]
^{1/2}}{2\pi (\rho \delta)^{1/2}\sigma(\beta_w)} d w\right\},
\end{equation}
where the function $\nu$ has the same argument
as in the preceding approximation, and
where $\Sigma (w) = \Expec\{-D^2_w [\ell(t, w,r) - x_{w,r}]\}.$
A proof of (\ref{generalapprox}) can be obtained from the
methods described in Section \ref{sec:tailapprox}.
An alternative approximation, which may be useful in some
special cases, is discussed below..

For maximization over
both $w$ and $r$, the integral becomes two-dimensional,
$\Sigma = \Sigma(w, r)$ is
the determinant of the expectation of the negative Hessian, and
there is one more factor of $1/(2 \pi)^{1/2}$.  A similar
result holds if there are
more parameters.  It appears that for the examples of this paper there are
significant edge effects when one maximizes over $r$, so the first order
asymptotic approximation given here may not be adequate, with resulting
implications for the power.  We return to this point below.

\begin{remark} \label{rem:chipseq}
Since the function $g$ in the definition of $Z_t$ is
a smooth function of $w$, the approximation (\ref{generalapprox})
contains the essential ingredients for an approximate p-value for the scan statistic (\ref{eq:chipseq}) used in ChIP-Seq analysis.
See also Section 6.5 below.
\end{remark}

\begin{remark}
When we fix the values of the nuisance parameters
$w,r$ the statistics, $ \max_t Z(t; w,r)$ and $\max_t \ell(\beta, w,r)$
are equivalent in the sense that a suitable threshold for one is a
known linear function of the corresponding threshold for the other.
This is not so if we maximize with respect to
$w$ or $w,r$.   See Table 2 below.
\end{remark}

\subsection{Special Cases}

There are two special cases of particular interest.
If the threshold $x$ does not depend on $w$ and $r$,  the equation defining
$\beta_w$ is $\rho \delta J(\beta_w,w,r) = x$, so (\ref{generalapprox})
specializes to
\[
1- \exp\{- m e^{-x}\rho\int_{w_0}^{w_1}[\xi(\beta_w)- \xi(0)]
\nu\{ \cdot \}
[\Sigma(w)]^{1/2}[(2\pi (\rho \delta)^{1/2}\sigma(\beta_w)]^{-1} d w\}.
\]
In this case
$\Sigma (w)$  is most easily computed as the variance of
$D_w\ell(t;w,r)$.

The second case of particular interest involves events of the form
$$\{ \max_{t, w_0 \leq w \leq w_1} Z(t;w,r) \geq x_0\},$$
which can be written in the form of (\ref{lrinequality}) by
putting $x_{w,r} = \beta x_0 - \rho \delta \psi(\beta;w,r).$
In this case $\beta$ is chosen to satisfy $x_0 = \Expec[Z(t,w,r)]$.
The Kullback-Leibler information $J$ in (\ref{generalapprox}) typically
has a minimum value
inside the range of integration over $w$, so the integral
can be approximated by Laplace's method.
This can be particularly useful in multi- dimensional problems,
where numerical integration can be onerous.

For example, suppose
that $w^*$ denotes the minimizing value of $w$.  It is shown
below that at the minimum the negative second derivative of $J$ with respect to
$w$ is  $$D^2_w J = - \beta^* \Expec_{\beta^*} [D^2_w Z(w^*)]
+ {\rm Var}_{\beta^*} [\beta^* D_w Z(w^*) | Z(w^*)],$$
where $\beta^* = \beta(w^*).$  Also
$$\Sigma(w^*) =  -\beta^* \Expec_{\beta^*} [D^2_w Z(w^*)]$$ and
$$D_w \psi = \beta^* \Expec_{\beta^*} [D_w Z(w^*)].$$
The integral in (\ref{generalapprox}) is asymptotically equal to
\begin{equation} \label{Laplace}
\exp[-\rho \delta J(\beta^*,w^*,q)]
[\xi(\beta^*)- \xi(0)]
\nu\{ \cdot \}
[\Sigma(w^*)]
^{1/2}/[2\pi \rho \delta \sigma^2(\beta^*)\{D^2_w J\}]^{1/2}.
\end{equation}

For a numerical example, for $x = 11.5, \delta = 200, m=1000000,
\rho = 0.5, r = 0.1, w_0 = 0.5, w_1 = 5$, (\ref{generalapprox})
yields
the value 0.053, while (\ref{Laplace}) gives 0.051.  For
$\rho = 1, \delta = 100$ and the other parameters unchanged,
(\ref{generalapprox}) gives 0.105, while (\ref{Laplace}) gives 0.101.

If the maximum with respect to a parameter occurs at an endpoint
of the interval of maximization, a somewhat different simplification of
(\ref{generalapprox}) is appropriate.  An example is the parameter
$r$, where the maximum occurs at the upper endpoint. In such cases
it may also be appropriate to add an edge correction.

\subsection{Approximation accuracy}

We have performed  a small Monte Carlo experiment to
evaluate the accuracy of (\ref{special}).  The number of
repetitions of the Monte Carlo experiment is 2000
(except for the last two rows, where the number of repetitions
was increased to 2500).
The threshold $x$ and sample size $m$ are
smaller than one might want to use
in practice, since a Monte Carlo experiment for more realistic
values would be extremely time consuming.  The table contains
two analytic approximations.  For the first we take $\nu = 1$,
while the second uses the computed value of $\nu$.  In almost all
cases the second approximation is more accurate although
usually still somewhat conservative.  To compare these
results with those that come later, it may be helpful to
imagine that the statistics are evaluated along a grid with
spacing of 10 base pairs.  Then the values of $\rho$ in units
of base pairs are the tabled values divided by 10, while the
values in base pairs of $\delta$,
$w$, and $\sigma$, are the tabled values multiplied by 10.
The first approximation to the p-value would remain unchanged,
while the second, which incorporates the step size in the statistic
would increase.

\begin{table}[h]                \caption{Approximate p-values}
\begin{center}
\begin{tabular}{c|c|c|c|c|c|c|c|c|c}
\hline
$m$ & $x$  & $\rho$ & $ \delta $ & $w$& $r$ & $\sigma$ &Approx1  & Approx2  &  Monte Carlo \\
\hline
1000 & 6.10  & 1 & 40 & 3.0 & 0.1 & 4 & 0.050 & 0.034 & 0.037 \\
1000 & 6.10 & 4 & 40 & 3.0 & 0.1 & 4 & 0.061 & 0.040 & 0.050 \\
1000 & 6.10 & 10 & 40 & 3.0 & 0.1 & 4 & 0.066 & 0.043 & 0.053 \\
1000 & 6.15 & 1 & 20 & 3.0 & 0.1 & 1 & 0.050 & 0.037 & 0.016 \\
1000 & 6.15 & 1 & 20 & 3.0 & 0.1 & 1 & 0.066 & 0.044 & 0.028 \\
2000 & 5.00  & 1 & 20  & 1.5 & 0.1 & 1 & 0.059 & 0.050 & 0.045\\
2000 &5.00 & 0.5 & 20 & 1.5 & 0.1 & 1 & 0.052 & 0.045 & 0.042  \\
2000 &5.00 & 0.25 & 20 & 1.5 & 0.1 & 1 & 0.045 & 0.040 & 0.027 \\
2000 & 5.00 & 0.5 & 20 & 2.0 & 0.2 & 1 & 0.056 & 0.048  & 0.044   \\
2000 & 4.00 & 0.25 & 40 & 3.0 & 0.1 & 1 & 0.053 & 0.049 & 0.039 \\
2000 & 4.00 & 0.25 & 40 & 3.0 & 0.2 & 1 & 0.062 & 0.057 & 0.053  \\
2000 & 4.00 & 0.5 & 40 & 3.0 & 0.1 & 1 & 0.061 & 0.055 & 0.039 \\
2000 & 5.15 & 0.5 & 20 & 1.5 & 0.2 & 1 & 0.052 & 0.044 & 0.040 \\
2000 & 4.00 & 0.25 & 20 & 3.0 & 0.1 & 1 &  0.090 & 0.081 & 0.047 \\
2000 &6.50 & 0.25 & 200 & 1.5 & 0.1 & 1 & 0.011 & 0.010 & 0.006 \\
2000 & 7.20 & 1 & 100 & 1.5 & 0.1 & 1 & 0.013 & 0.010 & 0.007  \\\hline
        \end{tabular}
        \label{table:DD}
\end{center}
\end{table}

The effect of using the negative binomial process is surprisingly
small.  For example, for the ninth row of Table 1,
using a gamma mixture of Poisson processes as suggested above
with $\alpha = 2$ produces
the same p-value (0.056) to two significant figures.  The much
smaller values $\alpha = 0.05$ and 0.025 produce the approximate p-values
0.065 and 0.070.

\subsection{Non-homogeneous processes}

In the case that the underlying Poisson process is
non-homogeneous with intensity $\rho(t)$,  the approximations given above apply with only slight modifications.  Consider
the case of fixed $r$.  Since the measure transformation (\ref{mt}) decomposes the boundary crossing probability into a sum of $m$ terms, each depending on $t$, the expressions given in the exponents in the
approximations (\ref{special}, \ref{generalapprox}) changes to a sum of $m$ terms, instead of a single expression multiplied by $m$.  For the $t$th term, the definition of $\beta=\beta_{t,w}$ depends
on both $t$ and $w$, since $\rho \delta$ in the definition of the
cumulant generating function and throughout the
approximation is replaced by $\Omega(t)$.
In addition the factor $\rho[\xi(\beta_{t,w}) - \xi(0)]$ becomes
$\rho(t) \xi(\beta_{t,w}) - \rho(t+\delta) \xi(0).$
%And the argument of $\nu$ now
%has (in addition to the preceding expression in the numerator) a denominator
%that equals $\rho(t) \sigma^2(\beta) + \rho(t+w) \sigma^2(0)$.
%Finally the expression $\rho w$ is replaced by $\int_t^{t+w} \rho(s) ds$.

The approximations given above involve a number of different moments, which
are easily calculated as one dimensional numerical integrals.  Appropriate
formulas are given below.

\subsection{Piecewise smooth processes}
Consider the model (\ref{chipseq}), where the process $Z_\tau$ varies
smoothly with $\tau$.  Assume that $Z_{\tau, w} = \int g_w(\tau - t) dN_t$, for
a twice
differentiable kernel $g$
and consider
$\Prob \{ \max_{\tau,w} Z(\tau,w) \geq  x \}$,
where the max extends over $\tau_0 < \tau < \tau_1$ and $w_0 < w < w_1$.
Then $\ell(\tau) = \beta Z(\tau, w) - \psi(\beta; \tau, w)$,
where $\psi(\beta; \tau, w) = \int\{\exp[\beta g_w(\tau - t)] -1\} \rho(t) dt$,
and $J(\beta;\tau,w) = - \int\{\exp[\beta g_w(\tau - t)] -1 - \beta g_w(\tau-t)\}
\rho(t) dt.$  Setting $\beta$ to satisfy $\Expec[Z(\tau, w)] = x,$ we find
that the probability of interest is approximately
\begin{equation} \label{continuousapprox}
\int_{\tau_0}^{\tau_1} \int_{w_0}^{w_1} \exp[- J(\beta; \tau,w)]
\Big[\frac{\Expec\{-D^2[Z(\tau,w) - x]\}}
{{\rm Var} \ell_\tau}\Big]^{1/2} dw d\tau/(2\pi)^{3/2},
\end{equation}

In the case $\rho(t) = \rho$ for all $t$, the integrand is
a constant function of $\tau$, except near the end-points,
so the integral with respect to $\tau$ can be simplified
to multiplication by $\tau_1 - \tau_0$.
See Siegmund and Worsley (1995) for justification and  examples in the
case of Gaussian processes.

Since the range $w_1 -w_0$ may not be large, it
may be helpful to include a boundary correction.  One possibility
is to add to (\ref{continuousapprox}) the probability for
a fixed value of $w_0$.

%Presumably the
%integrand is a decreasing function of $w$, so one could
%also simplify that integration by an asymptotic approximation
%similar to that used to obtain (\ref{Laplace}), but now
%with the maximum of the integrand appearing at an endpoint of
%the interval of integration.

\subsection{An alternative approximation in a special case}

In the special case that there is only one ``smooth'' parameter over which
we maximize,
e.g., $w$ in the toy model for paired end reads or $\tau$ in the
score statistic for the model (\ref{chipseq}) with $w$ fixed, one can use
a derivation arising from an upcrossing (or downcrossing) argument
to give a formally different, although related, approximation.
To simplify the following brief calculation, suppose we
Let $Z_\tau = \int g(\tau - t) dN_t$, for a twice
differentiable kernel $g$
and consider
$\Prob \{ Z(\tau) \geq x(\tau)\; {\rm for \; some} \; \tau_0 \leq \tau \leq \tau_1 \}.$
Then $\ell(\tau) = \beta Z(\tau) - \psi(\beta; \tau)$,
where $\psi(\beta; \tau) = \int\{\exp[\beta g(\tau - t)] -1\} \rho(t) dt$,
and $J(\beta;\tau) = - \int\{\exp[\beta g(\tau - t)] -1 - \beta g(\tau-t)\}
\rho(t) dt.$

For notational convenience let $\tau_0 = 0$.
Partition the interval $[0, \tau_1]$ at points equally spaced at
distance $\Delta$, which we will let converge to 0.
Then as an upper bound for the corresponding ``discrete time'' maximum,
we have, up to a boundary term
(which may or may not be important),
\begin{equation} \label{discreteRice}
\sum_i \int_0^{\infty} \Prob_0 \{Z_{i\Delta} \in x(i\Delta) + \Delta dy,
Z_{(i+1)\Delta} < x((i+1)\Delta)\}.
\end{equation}
After an exponential family change of measure determined by choosing $\beta
= \beta_\tau$ so
that under the formal alternative $\Expec(Z_{\tau}) = x(\tau)$
and some minor algebraic manipulation, the $i$th term becomes
\begin{eqnarray} \label{tiltRice}
\int_0^ \infty &&\exp[-J(\beta,i\Delta)] \exp(-\beta \Delta y)\\ &\times& \Prob\{ Z_{i\Delta} \in x(i\Delta) + \Delta dy \}\Prob \{
[Z_{(i+1)\Delta} - Z_{i\Delta}]/\Delta < -y | Z_{i \Delta}\}. \nonumber
\end{eqnarray}
Under the formal alternative $Z_{\tau}, Z'_{\tau}$ are jointly asymptotically
normal.  Letting $\Delta \rightarrow 0$ and employing the asymptotic normality,
we find after simple manipulations that we get as an asymptotic
approximation for (\ref{discreteRice}) the expression
\begin{equation} \label{Rice}
\int_{\tau_0}^{\tau_1} \exp[-J(\beta; \tau)] \Big[\frac{{\rm Var}(Z'_\tau|Z_\tau)} {
2 \pi {\rm Var} Z_\tau}\Big]^{1/2}\eta(\xi)  d\tau,
\end{equation}
where $\eta(y) = \phi(y) + y \Phi(y)$ and
$\xi = [x'(\tau) - \Expec(Z'_\tau)]/[{\rm Var}(Z'_\tau|Z_\tau)]^{1/2}.$

It is natural to  ask how this approximation compares with
\begin{equation} \label{pws2}
\int_{\tau_0}^{\tau_1} \exp[- J(\beta; \tau)] \Big[\frac{\Expec(-D^2_\tau(\ell_\tau
 - x_\tau)} {
{\rm Var} \ell_\tau}\Big]^{1/2} d\tau/(2 \pi),
\end{equation}
obtained as in the preceding section.
In general the two results are
not the same and numerically yield slightly different approximations.
But if
$x(\tau) = [x + \psi(\beta; \tau)]/\beta$, so that $Z_\tau > x(\tau)$
if and only
if $\ell(\tau) > x$ and $J(\beta; \tau) = x$, it may be shown
that the two expressions are algebraically identical.
In fact, the preceding derivation may be carried out with $\ell(\tau)$
in place of $Z(\tau)$.  Then differentiating the equations
$1 = \Expec_0[\exp(\ell(\tau))]$ and $x = \Expec_0[\ell(\tau)\exp(\ell(\tau))]$,
we see that $\Expec[D_\tau \ell(\tau)] = 0 = \Expec[\ell(\tau) \ell'(\tau)]$,
so $\Expec[-D^2_\tau \ell(\tau)] = {\rm Var}\{D_\tau[\ell(\tau)] |\ell(\tau)\}$,
from which identity of the two expressions is easily derived.

It is also possible to derive a discrete time upper bound starting
from (\ref{discreteRice}), which while expected to be conservative
has the advantage that it can be applied to either smooth or not
smooth processes or a combination of both.
To simplify matters slightly
assume that $x({i \Delta}) = x_1$ for all $i$.
By likelihood ratio arguments, one can see that
$\Expec(Z_{t+\Delta}|Z_t) = \rho_0 Z_t$
and ${\rm Var}(Z_{t+\Delta}|Z_t) =\sigma_0^2 (1-\rho_0^2) $, where
$\sigma_0^2 = {\rm Var}_0 (Z_t)$ and $\rho_0 = {\rm Corr}_0 (Z_t, Z_{t + \Delta})$.
Using a normal approximation and these conditional moments, we replace
the conditional probability in (\ref{tiltRice})
by $\Phi\{ -[\rho_0 \Delta y - (1-\rho_0) x_1]/\sigma_0 (1-\rho_0^2)^{1/2} \}$
and the marginal probability by $\Delta dy/(2 \pi \sigma^2_\beta)^{1/2}$, then integrate
to get an approximate upper bound for the term indexed by $i$.

%However, in the important case of the score statistic, so $x$ is fixed
%and does not depend on $\tau$, we can again obtain a simplified approximation
%along the lines of (\ref{Laplace}).  In (\ref{Rice}) the argument of $\eta$
%at the maximizing value of $\tau$ is 0, so the value of $\eta$ is
%$1/(2\pi)^{1/2}$, and hence the asymptotic approximation to (\ref{Rice})
%is the same as before.

\subsection{Marginal Power} \label{sec:power}

The detection statistics of the preceding section are all of the
form $\max_\xi Y_\xi$.  Suppose that under some suitable model
$\Prob \{ Y_\xi \geq x\}$ is maximized at $\xi = \xi_0$.
It seems reasonable to define the local power of the detection
scheme to
\begin{equation} \label{locpow}
\Prob\{ Y_{\xi_0} \geq x \} +
\Prob \{ Y_{\xi_0} < x, \max_\xi Y_\xi \geq x\}.
\end{equation}
Since the second term is usually very small compared to the first
in cases of interest, we define $\Prob\{ Y_{\xi_0} \geq x \}$
to be the marginal power.
In this section we consider again a homogeneous process and use
the marginal power, evaluated by means of a
normal approximation,
to compare different procedures.

For the following example we assume that $\delta = 200,$
$m = 1000000,$ $\rho = 0.5$ and maximize over
$[0.5  <  w  \leq 5]$.  The marginal power is given
for four different statistics:  $Z = \max_{t, w} Z(t; w, r_0)$ and
$\ell = \max_{t,w} \ell(t; w, r_0)$, where $r_0 = 0.1$, for
$\ell_2 = \max_{t, w, r}
\ell(t; w, r)$, where the maximum over $r$ is restricted to the
range $0.03 \leq r \leq 0.2$, and finally $\ell(w_0,w_1; 0.1)
= \max [\ell(w_0,0.1)/b_0, \ell(w_1,0.1)/b_1],$ where $w_0 = 1.0, w_1 = 3.5.$

We have assumed that the standard deviation of $F_0$ is one,
whereas in at least some applications it is about 10.
Then by a rescaling argument a shift in the distributions of the amount
$w$ corresponds to $10 w$  base pairs on a genomic scale.
In examples with longer fragments the standard deviation appears to be
about 60, so $w$ corresponds to $60 w$ base pairs.

For all statistics the significance level based on the approximations
given above with $\nu = 1$ is about  0.05.
%To determine
%suitable thresholds we used the approximations given above.
%since $\rho$ is small, we set $\nu(\cdot) = 1$, which is
%an upper bound under completely general conditions.
For $\ell(2;0.1)$, $Z$ and $\ell$, we obtained
the thresholds $x = 11.4$,  11.54 and 12.05, respectively.  For
$\ell_2$, the situation is more complicated, since the tail
probability is largest at the largest values of $r$.  Hence
as an approximation for the significance level,
as an edge correction to the approximations
involving the max over $w, r$ given above, we have added the max over $w$ at $r_1$, the
maximum value of $r$.  This produced the threshold 12.87.
For $\ell(w_0, w_1, 0.1)$ we used a Bonferroni bound to combine the two
statistics, where
$b_0 = 12.34$ and $b_1 = 11.9$ were chosen so that
the individual statistics had
0.025 significance level.  The column headed ``Opt'' gives the power
for the statistic $\max_t Z(t;w,r)$ for the indicated values of
$w,r$ and the 0.05 threshold (which also depends on $w,r$ and
is omitted.)   Although we do not know whether using the true
parameters to define the log likelihood ratio actually
achieves maximum power, it seems a reasonable
measure of the power that might be achieved with complete knowledge of the
parameters.  The statistic $\ell$, which uses $r_0 = 0.1$ and is adaptive
with respect to $w$ does remarkably well.

\begin{table}[h]
\caption{Parameters are  $\rho = 0.5, \;r = 0.1, \; \delta = 200, \;
m = 1000000; \;\; \max$ over
$w \in [0.5,5]$,
$r \in [0.03,.3].$
}
\begin{center}
\begin{tabular}{c|c|c|c|c|c|c|c}
\hline
 $r_1$  & $w_1$ & ``Opt'' & $\ell(2,0.1)$ &$ Z(0.1)$& $\ell(1,3.5; 0.1)$  & $\ell$  &  $\ell_2$ \\
\hline
0.1  & 2.5 & 0.53 &0.52& 0.52 & 0.47 & 0.50 & 0.45 \\
0.1 & 3.0  & 0.82&0.80& 0.81  & 0.80 & 0.80  &  0.78  \\
0.1   & 2.25 & 0.32&0.32 & 0.31 & 0.21 & 0.29 & 0.24 \\\hline
0.3  & 1.4 & 0.54 &0.43&0.38 & 0.47 & 0.50 & 0.46 \\
0.3 & 2.0 & 0.96& 0.96 & 0.95 & 0.95& 0.96 & 0.96 \\\hline
0.5 & 1.0 & 0.63 &0.31& 0.26 & 0.48 & 0.59 & 0.54 \\
0.5 & 1.5 & 0.99 & 0.96& 0.95 & 0.98 & 0.98 & 0.98 \\\hline
0.03 & 4.0 & 0.55& 0.37& 0.43 &0.46 & 0.50 & 0.47 \\
0.03 & 4.5 & 0.70&0.51 & 0.58 &0.61 & 0.66 & 0.64    \\\hline
0.02 & 5.0 & 0.64&0.38 & 0.48 &0.50 & 0.58 & 0.55   \\\hline
        \end{tabular}
        \label{table:DD}
\end{center}
\end{table}

From these numbers it appears that when $r$ is  not too far from the
assumed value, $r_0$, the statistic $Z$ is slightly more powerful
than $\ell$, but it can be considerably less powerful when
$r$ is quite different from $r_0$.  The statistic $\ell_2$
seems less powerful than $\ell$, even when the actual value of
$r$ is not close to the assumed value $r_0$. It is possible that
the performance of $\ell_2$ has been adversely affected by our
{\sl ad hoc} method of controling the significance level.
The statistic $\ell(1,3.5;0.1)$ is much simpler
than $\ell$ and seems to be only slightly, but consistently less powerful
over the range of parameters considered here.

%The range over which $w$ was maximized, $[0.5,5]$ was chosen somewhat
%arbitrarily, and particularly for larger values of $\rho$ it may be
%possible to detect smaller values of $\mu4.$  Hence it may be useful to note
%that the value 0.5 can be decreased at the cost of only  modest increases in the significance level,
%and hence the required threshold.

The rate parameter $\rho$ of the driving Poisson process is
effectively the sample size, hence an
important determinant of the power.  Smaller values of $\rho$
lead to lower significance thresholds but, evenso, to less power.
A natural question concerns the extent to which the loss of
power associated with a smaller value of $\rho$ might be
mitigated by using a lower threshold.
For a simple numerical example, for the scenario described above
but with $\rho = 0.2$, the statistic
$\ell$ would have the significance threshold
approximately  $x = 11.54$.  The marginal power
at $r = 0.1, w = 3.0$, which was approximately 0.8 when
$\rho = 0.5$,  would now be about 0.34.  If we do not
use the appropriate 0.05  threshold, but instead use the
original threshold, $x = 12.05$ appropriate for $\rho = 0.5$,
the power would be about 0.32---only
slightly less.    For another example, suppose $r = 0.5, w = 1.5$.
The marginal power for the ``correct'' level 0.05 threshold of 11.54
would be
0.57, while for the original threshold of 12.05 it would be 0.54.
These numbers and more extensive calculations not reported here
indicate that the loss of power from a moderately smaller value of $\rho$
is intrinsic and cannot be
compensated by using the smaller threshold appopriate for the smaller value
of $\rho$.
Consequently, in the case of a non-homogeneous process,
it seems very difficult to gain power by trying to vary the
threshold locally to accommodate changing values of $\rho(t)$, at least
when the range of variation of $\rho(t)$ is not too large.

\section{Analysis of Scan Statistics for Structural Variants} \label{sec:sv2}

We now consider the more detailed model to detect insertions and deletions proposed in
Section \ref{sec:bettermodel}.  The log-likelihood ratio statistic under this model
is a sum of the signature-specific scores.  In practice, each score can be used on its own as a scan statistic, or they can be summed in various combinations.
Our power comparisons below show that the different scores achieve power in
different regions of the parameter space.  Although summing them improves power
under specific conditions, overall it does not significantly improve power
compared to  applying each score individually and then adjusting the p-value by
the Bonferroni inequality.  In addition, tail probability approximations
for the summed
statistic involves more tedious derivations which are, in effect, a combination of the
terms for the individual scores.  Given these considerations, we will discuss
specifically control of the false positive rate only for the individual
scores $Z^B_t$ and the sum of the hanging read scores,  $Z^+ +  Z^-$.

The parameters $w,r$ are given nominal values, except for hanging reads
in the case of deletions, where we maximize over a range of $w$ as
discussed above to align peaks.

Consider first the score $Z^B$ for detecting deletions
using bracketing pairs.
Here, the parameter $\tau$ is the triple $(s,w,r)$.  The kernel function
corresponding to the alternative is $k_{\tau}(z) = \log[1-r+rf(v-u+w)/f(v-u)]I(z \in S^B_{s,w})$.  It will be convenient to put $g(x) = g(x;w,r) = \log[1-r+rf(v-u+w)/f(v-u)]$,
so the cumulant generating function of $Z^B_{\tau}$ is given by
\begin{equation} \label{psiZB}
\psi_{\tau}(\beta) = (1-p)\int_{u<s-R} \kappa(u) \int_{v>s+w} \kappa(v) f(v-u)
\{\exp[\beta g(v-u)]-1\}dvdu.
\end{equation}
In the case of a homogeneous process this simplifies to
$(1-p)\kappa^2 \psi_1(\beta)$, where
\begin{equation} \label{psi1ZB}
\psi_1(\beta) = \int_{w+R}^{\infty} (x-w-R)f(x)\{\exp[\beta g(x)]-1\}dx.
\end{equation}
A similar analysis applies to insertions, in which case $w$ is the negative of the insert
size, the range of integration for $v$ in (\ref{psiZB}) changes to $v>s$, and
the range of integration in (\ref{psi1ZB}) changes to $(R,\infty)$.    Given the
cumulant generating function, the false positive rate for a scan using $Z^B_{\tau}$ can be obtained along the lines of the results for the mixture model.  In particular,
for fixed $w,r$ we have the approximation (\ref{special}) with $\delta=1$,
since $\delta$ is incorporated into the definition of $\psi_1$.  The calculations of the
parameter $\xi(\beta)-\xi(0)$ are more complicated.  Some details are sketched
in an appendix.

Numerical examples for scans using $Z^B_{\tau}$, with $R = 36,
\; p = 0.03, \; \delta = 200, \;
\sigma = 10$, indicate that the statistics behave similarly to those discussed for
the toy mixture model, although the power of a scan under the more precise model is somewhat more for both insertions and deletions, and is somewhat larger for deletions than for insertions.
One distinction worth noting between insertions and deletions in this model is that
while power increases with the length of a deletion, it can decrease for insertions when $w$  becomes a
substantial fraction of the insert length, since an insert must span the insertion in
the target genome for the read pair to be informative.

Now consider the scores $Z^+$ and $Z^-$, or their sum,
which uses hanging reads for detection.
%For example, the kernel function for $Z^-$ is $k_{\tau}(z) = \log(1+2rp^{-1}(1-p)\{F(s-u)-F(s-R-u)\})I(z \in S^-_{s,w})$.
Since these scores have piecewise
smooth sample paths, we can use the approximations (\ref{Rice}) or (\ref{pws2}),
modified in the case of deletions to account for a second maximization
as described in Remark \ref{rem:alignpeaks}.

In what follows we consider what we find after some numerical experimentation to
be reasonable fixed values of the parameters $w,r$.  It is also possible to
maximize $Z^B$ over $w,r$ and $Z^+$, $Z^-$ over $r$.  This would require changes
to the approximation in fashion similar to (\ref{generalapprox}), but more complicated.

\subsection{\bf Power comparison} \label{sec:power2}

We now examine the power of the tests to detect insertions and deletions based on the
bracketing pair score $Z^B$ and the hanging pair score
$Z^H = Z^++Z^-$ (maximized over $0 \leq w \leq 150$ for deletions,
as explained above).  Using a normal approximation,
the marginal power can be easily computed as described in Section \ref{sec:power}.

Insertions and deletions are considered separately.  The sequencing and
library preparation parameters that influence power are the length of the read, $R$,
the mean $\delta$ and standard deviation $\sigma$ of the insert length distribution,
and the sequencing coverage (that is, the average value of $R\kappa^2$).  Power of the hanging reads score also depends heavily on the value of $p$, the probability
of a mapping error that leads to a hanging read under the null hypothesis. Together,
these parameters determine  the null distribution.

Table \ref{table:RealData} shows the value of $R$ and the estimated values
of $\delta$, $\sigma$, and $p$ for a few typical publicly available data sets.
The first three data sets in the table are samples sequenced to high depth by the 1000 Genomes Consortium.  The last two are samples sequenced by Illumina Corporation
as part of their Platinum Genomes initiative, the goal of which is to provide a
set of high quality ``gold standard'' sequencing data for testing and validation of
different methods.  The first sample was produced in 2011, when the standard read
length was 36 and shorter fragment lengths (mean 197) were the norm.  The last four
samples are more recent, and reflect the trend towards increased read
and fragment
lengths. With shorter reads ($R=36$) many repetitive regions of the genome can not be
mapped, leading to an
estimate of $p$ to be between 0.01 and 0.05.
In one set of data with $R=100$
$p$ appears to be about $0.033$.  Note that increased fragment
lengths come at a cost of an increased standard deviation, and thus the
effect of
this more recent protocol on power is not so clear.  Based on this table,
we will
analyze power under a number of assumptions about the null parameters
with emphasis on two settings that we have observed in data:
$R = 36,$ $p = 0.03 $,
$\delta = 200$,  $\sigma = 10$ and $R = 100$, $p = 0.033$, $\delta = 220$,
$\sigma = 63$.  We also consider a few examples with longer insert lengths
and smaller $p$.

Although the values of $R$, $p$, $\delta$ and $\sigma$ more or less fall within
standard ranges for sequencing studies conducted during the same time period,
coverage can vary widely across studies, and depends on the goals of the experiment
and how much the investigator wants to invest in the experiment.
Currently, ``low-coverage'' usually refers to cases where each genomic position is
covered by an average of 10 reads or less, and ``high coverage'' to cases where each
genomic position is covered by an average of 40 reads or more.  In some studies,
for example in the study of evolving virus populations or circulating tumor DNA,
extremely high coverage in the hundreds or thousands, is desired.  These are referred
to as ``deep sequencing'' experiments, where the mutations of interest are sometimes
present at very low frequencies ($r<1\%$) in the sample.

We will examine two scenarios for coverage.  For the first  $\kappa^2 = 0.27$,
which we observed in data and
represents moderate coverage.
For the second  $\kappa^2=5$ and we study only the $R=100$ setting,
which represents deep sequencing  with an average coverage of 500.   In both cases,
we let $m = 1000000$.  Larger values of $m$ are likely to occur in
practice, but do not seem to yield additional insights.

We found through simulations and numerical studies that, as for the toy mixture model,
the power of the scores is not particularly sensitive to the \emph{assumed}
values of $r$ and $w$ used to define the scores.  For simplicity, we set
$w = 30$ base
pairs in $Z^H$ for insertion, and $|w| = 30$ base pairs for $Z^B$.  The assumed value
of $r$ is set to 0.1 in all statistics.

For the most part, power increases with the true size of the insertion/deletion
($w$) and its true frequency in
the sample ($r$), both of which are properties of the alternative distribution.
These are chosen so that at least one of $Z^B$ and $Z^H$ has moderate power.  We expect
to find that $Z^H$ has relatively more power than $Z^B$ to detect short variants, and relatively less to detect longer variants.

\begin{table}[h]     \caption{Features of Several Public Data Sets}
\begin{center}
\begin{tabular}{l|l|l|c|c|c|c}
\hline
 Source & Sample name & Date & $R$ & $\hat{\delta}$ & $\hat{\sigma}$  & $\hat{p}$ \\
\hline
1000 Genomes & NA12878 & Nov, 2011 & 36 & 197 & 9.6 & 0.01-0.05\\
1000 Genomes & NA12878 & July, 2013 & 100 & 398 & 33 & 0.01-0.05\\
1000 Genomes & NA12891 & July, 2013 & 100 & 342 & 70 & 0.01-0.05\\
Illumina Platinum Genomes & NA12878 & July, 2013 & 100 & 220 & 63 & 0.033 \\
\hline
\end{tabular}
        \label{table:RealData}
\end{center}
\end{table}

Consider first the case of insertions under moderate coverage ($\kappa^2=0.27$).
For the case of $R=36, \delta = 200, \sigma=10,p=0.03$ the 0.05 threshold
for $Z^H$ is  $12.3$.  The corresponding threshold for the the statistic based on $Z^B$ is
$10.4$.  For $R= 100, \delta = 220, \sigma = 63,p=0.033$, the threshold
for the statistic based on $Z^H$ is $x_1 = 15.2$.  For the statistic based on $Z^B$ it is
$x_1 = 0.21$.  Marginal power for varying values of $(r,w)$
is given in Table \ref{table:II}.  Observe that when $R=36$, $Z^B$ has
better power than $Z^H$ when the insertion size is large, provided that it
is still somewhat more than the length of the insert.  The
statistic $Z^H$ has better power than $Z^B$ to detect short indels at
high frequency.  When $R=100$, $Z^B$ has no power in the
situations studied here, except for the case where the mean insert length
was 400.

\begin{table}[h]     \caption{Marginal Power:  Insertions}
\begin{center}
\begin{tabular}{c|c|c|c|c}
\hline
 $R,\delta,\sigma,p$ & r & w & Hanging Reads  & Bracketing Pairs  \\
\hline
36,200,10,0.03 & 0.5 & 10  & 0.86 & 0.00    \\
36,200,10,0.03 & 0.5  & 20  & 0.95 & 0.74 \\
36,200,10,0.03 & 0.5 & 100  & 0.99 & 1.00 \\
36,200,10,0.03 & 0.2 & 50 & 0.32 & 0.71 \\
36,200,10,0.03 & 0.1 & 100 & 0.03 & 0.80 \\
36,200,10,0.03 & 0.1 & 150 & 0.03 & 0.51\\
100,220,63,0.033 & 0.5 &10 & 1.00 &  0.00 \\
100,220,63,0.033 & 0.5 & 100 & 1.00 & 0.00  \\
100,220,63,0.033 & 0.1 & 100 & 0.19 & 0.00 \\
100,220,63,0.033 & 0.1 & 200 & 0.28 & 0.00\\
100,220,63,0.01 & 0.1& 200 &  0.79 & 0.00 \\
100,220,63,0.033 & 0.2 & 10 & 0.41 & 0.00 \\
100,220,63,0.01 & 0.2 & 10 & 0.85 & 0.00 \\
100,220,63,0.033 & 0.2 & 100 & 0.88 & 0.00\\
100,400,63,0.033 & 0.5 & 100 & 1.00 & 0.72 \\
100,400,63,0.033 & 0.3 & 200 & 1.00 & 0.21 \\\hline
        \end{tabular}
        \label{table:II}
\end{center}
\end{table}

Table \ref{table:DD} shows the power for detecting deletions of
varying $(r,w)$ when coverage is moderate.  Unlike for insertions, the power
of $Z^B$ for deletions increases monotonically with the size of the deletion,
because deletions have width 0 in the target sample and thus can always be
captured within a bracketing fragment.  In comparison to $Z^H$, $Z^B$ has
better power when deletion size is large, and when $r$ is small.
As in the case of insertions, $Z^H$ has better power for longer reads
whether $p$ is smaller or not,
and it is preferable to $Z^B$ when r is large.  The power
of $Z^H$ does not depend on the size of the deleted region.
Perhaps surprisingly, in cases where the bracketing statistic has adequate
power, it has substantially more power when the short read/short fragments
are used, since the smaller standard deviation more than compensates.
See, for example, the rows with $r = 0.1$, $w = 100$ or those with
$r = 0.05$ and $w = 250.$

\begin{table}[h]     \caption{Marginal Power:  Detecting Deletions with
$\kappa^2 = 0.27$, $m = 10^6$ }
\begin{center}
\begin{tabular}{c|c|c|c|c}
\hline
$R,\delta,\sigma,p$ & $r$ & $w$ & Hanging Reads  &  Bracketing Pairs  \\
\hline
36,200,10,0.03 & 0.5  & 10  & 0.62 & 0.00    \\
36,200,10,0.03 & 0.5  & 20  & 0.62 & 0.84 \\
36,200,10,0.03 & 0.5 & 100  & 0.62 & 1.00 \\
36,200,10,0.03 & 0.1 & 100 & 0.00 & 0.99 \\
36,200,10,0.03 & 0.1 & 150 & 0.00 & 0.99\\
36,200,10,0.03 & 0.05 & 150 &0.00  & 0.96   \\
36,200,10,0.03 & 0.01 & 150 &0.00 & 0.64 \\
36,200,10,0.03,& 0.01 & 250 & 0.00 & 0.75\\
100,220,63,0.033 & 0.5 &10 & 0.99 & 0.00 \\
100,220,63,0.033 & 0.3 & 10 & 0.71 & 0.00 \\
100,220,63,0.033 & 0.3 & 100 & 0.71 & 0.40 \\
100,220,63,0.033 & 0.3 & 150 & 0.71 & 0.95 \\
100,220,63,0.033 & 0.2 & 150 & 0.25 & 0.75 \\
100,400,63,0.033 & 0.2 & 100 & 0.25 & 0.35 \\
100,400,63,0.01 & 0.2 & 100 & 0.75 & 0.36 \\
100,400,63,0.01 & 0.2 & 150 & 0.75 & 0.94 \\\hline
        \end{tabular}
        \label{table:DD}
\end{center}
\end{table}

%For both the detection of insertions and deletions under moderate coverage, we
%found that combining the bracketing pairs score and the hanging reads score
%by summing, i.e. $Z^H+Z^B$ rarely helps.  For most alternatives, one
%statistic usually dominates the other, and summing the two often leads to
%a  decrease in power.  Only selected  small regions of the
%parameter space does the summed statistic give an improvement in power.

Finally, we consider the case of detecting low-frequency mutations using deep
sequencing ($R=100, \delta = 400, \sigma = 63,p=0.033,\kappa^2=5$).  We
consider only deletions and examine the setting where the length of
the deletion is large and the frequency is small.  The 0.05 thresholds
for $Z^B$ and $Z^H$  are respectively 21.8 and 10.0.
Table \ref{table:DDD} shows the marginal power for varying $(r,w)$.
Compared to the scenarios in \ref{table:DD}, we see that $Z^B$ is more
competitive against $Z^H$ in the high depth, low $r$, large $w$ scenario.
%In several of the $(r,w)$ combinations listed, the combined statistic
%$Z^B+Z^H$ gives a significant improvement over using either statistic alone,
%although we must admit that these combinations were found after
%much searching.
%If $w$ were larger, the bracketing pairs score dominates
%hanging reads and combined scans, whereas if $r$ were larger, the hanging reads score dominates the other two scans.  However, in all the cases in
%Table 5, the short read/short insert bracketing pairs statistic has
%power at least 0.98, even though the power for hanging reads is
%essentially zero.

\begin{table}[h]     \caption{Marginal Power: Detecting Deletions with $R=100, \delta = 400, \sigma = 63,p=0.033,\kappa^2=5$ }
\begin{center}
\begin{tabular}{c|c|c|c|c}
\hline
$r$ &$w$ & Hanging Reads  &  Bracketing Pairs \\
\hline
0.10 & 5   & 1.00 & 0.00 \\
0.07 & 50 & 0.96  & 0.02\\
0.07 & 100 & 0.96 & 0.97 \\
0.05 & 150  &0.58 &1.00\\
0.02 & 200 & 0.00 &0.93\\
0.01 & 250  &0.00 &0.69\\\hline
        \end{tabular}
        \label{table:DDD}
\end{center}
\end{table}

\begin{remark} In our discussion of the power to
detect insertions and deletions,
we have concentrated on the ``marginal power,'' i.e., the first term in
(\ref{locpow}), which makes the major contribution to the overall power except
in a few cases where the power is itself small.  In some of
our examples, it is relatively easy to compute the generally more complicated
second term, and hence to see how much it contributes.  An illustration would
be an evaluation in Table \ref{table:II} of the row having $R = 36$, $r = 0.2$ and
$w = 30$, for which the marginal power of the bracketing pairs
statistic is 0.71.  Adding
an approximation for the second term in (\ref{locpow}) would bring the power
up to 0.75.  To compute this approximation we assume that
the process is Gaussian, which seems reasonable since power involves primarily
the center of the distribution, not the extreme tails, and adapt
the method of approximation of Feingold, Brown and Siegmund (1993).
For the same row, where the marginal
power for the hanging read statistic is 0.32, adding an approximation
for the second term in (\ref{locpow}) would produce an approximation of 0.38.
Here the method is based on the approximation found in
Siegmund and Worsley (1995).
\end{remark}

\begin{remark} In the case described in the preceding example, where both statistics
have some power, it is also possible to add the two statistics,
which leads to an increase in the approximate marginal power to 0.88.
For the row in Table \ref{table:DD} where $R= 100 = w$ and $r = 0.3$, again both
statistics have some power (0.72 for hanging reads, 0.40 for
bracketing pairs), adding the two statistics improves the marginal power
to 0.84.
The difficulty with this approach as a general strategy is that the
cases where it leads to an increase in power seem to require special
combinations of unknown parameters.  In most of the cases in
Tables 4-6  one of the statistics
dominates the other to such an extent that adding the two leads to a
loss of power, which can be substantial.
An alternative would be to combine the two statistics by taking
their maximum and making a Bonferroni adjustment to the significance
level.  If we take higher thresholds to make the individual significance levels
0.025, for the hanging read statistic the power would fall from 0.72 to 0.68,
while for the bracketing pairs statistic it would fall from 0.40 to
0.36.  Since these two statistics use reads from different genomic
regions, they are independent, and hence the max of the
two would have power 0.80.
\end{remark}

\subsection{\bf Data Examples} \label{sec:data}
\begin{figure}
\begin{center}
\includegraphics[scale=.45]{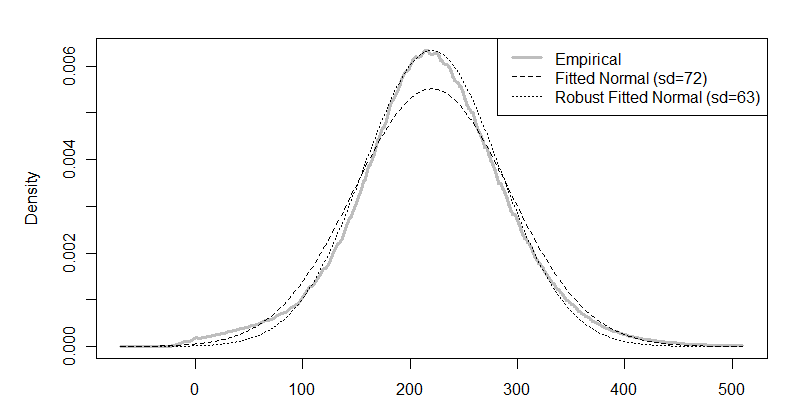} \\
\end{center}
\caption{\label{fig:ILdist} Empirical distribution of mapped insert sizes. }
\end{figure}

As an illustration, we consider a scan for deletions in the genome of
individual NA12878, which was sequenced by Illumina as part of the Platinum
Genomes project (Study accession PRJEB3381, Run accession ERR194147 in the European Nucleotide Archive).  We will limit our discussion to Chromosome 20, which has a total length of 63 megabases.  A total of 16,880,535 read pairs have at least one read mapped to this chromosome, which corresponds to a rate of 0.27 read-pair per base ($\kappa$=0.52).  The read length is 100 for these data. The mapped insert lengths have an empirical mean of $220$.  Figure \ref{fig:ILdist} shows the empirical insert length distribution, with the bold gray line showing a kernel density estimate, the dashed line showing the normal density with maximum likelihood estimates for mean and variance, and the dotted line showing the normal density with a robust estimate of variance.  The maximum likelihood estimate of standard deviation is 72, and the robust estimate is 63.  In practice, this difference in standard deviation does not make a big difference in the thresholds:  For a scan of the entire chromosome using $Z^B$ with parameters $w=20,~r=0.1$
and a step size of 10 bases, controlling the family-wise error rate at $\alpha=0.1$ leads to a threshold is 1.62 when assuming a standard deviation of 72, and 1.79 when assuming a standard deviation of 63.  The threshold for $\alpha=0.01$ is 1.74 for standard deviation of 72, and 1.94 for standard deviation of 63.

The proportion of hanging reads for these data is 3.3\%, of which about 2\% comes from pairs where one read is unmapped.  The $\alpha=0.1$ threshold for $Z^H$, using $r=0.1$ and a stepsize of 10, is 10.26.  The $\alpha=0.01$ threshold is 11.21

\begin{figure}
\begin{center}
\includegraphics[scale=.45]{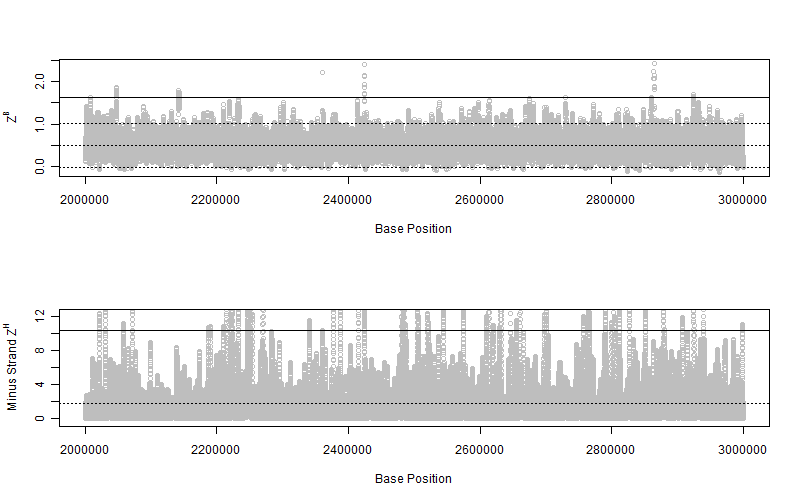} \\
\end{center}
\caption{\label{fig:data2} Insert length score $Z^B$ (top) and Minus strand hanging reads score $Z^-$ (bottom) scores for one megabase block on chromosome 20 of individual NA12878. The plus strand hanging reads score looks similar to $Z^-$ and is not shown.}
\end{figure}

It is difficult to visualize such a massive data set.  Figure \ref{fig:data2} shows the scores $Z^B$ and $Z^H$ for a quite typical one megabase long region.  Even at this resolution, the data are a blur. Overlayed on the plot for $Z^B$ are dashed lines, which represent the mean $+/-$ 3 standard deviations for the null distribution of the scores, which are computed analytically using our model.  For $Z^H$ only the mean is shown, since this score is heavily skewed. Note that most of the $Z^B$ process lie within this band, and that the null mean for $Z^H$ does seem to be at the right place.  Such visual checks give reassurance that the null model is a good approximation to the bulk of the data, and are an important part of the analysis.    The solid lines in the plots represent the threshold for family-wise error rate of 0.1.

Table \ref{tab:NA12878calls} shows the number of calls made by the insert length statistic $Z^B$ and the hanging reads statistic ($Z^+$ or $Z^-$) on Chromosome 20, and the number of places where the calls seem to be supporting the same variant.  For each score, overlapping windows where the score exceeds the threshold are merged into the same call.  For $\alpha=0.1$, 634 calls were made by $Z^B$, 3211 were made by $Z^+$ and 2790 were made by $Z^-$.  For $\alpha=0.01$, 399 calls were made by $Z^B$, 2935 were made by $Z^+$ and 2461 were made by $Z^-$.  At each p-value, about 5-6 times more calls were made by the hanging reads statistics. This may be due to a higher number of false positives due to lack of robustness, a higher sensitivity of the hanging reads statistic for insertions and small deletions, or a combination of these factors.  Without biological validation, it is hard to know.  We expect a true deletion to generate a peak in $Z^B$, coupled with minus strand hanging reads at the left boundary and plus strand hanging reads at the right boundary.  The number of regions called by $Z^B$ that overlap a call by $Z^-$ at the left end and a call by $Z^+$ at the right end is 58 for  $\alpha=10\%$ and 47 for $\alpha=1\%$.  This implies that only $10\%$ of calls made by the insert length statistic are supported by evidence from hanging reads.  Although the small overlap is cause for concern, visual inspection of the calls made by $Z^B$ that were not supported by hanging reads suggest that many of these calls may be real; an example is shown below.

\begin{table}
\begin{center}
\begin{tabular}{l|cc}
    \hline
  Method & $\alpha=10\%$ $\alpha=1\%$\\
  \hline
  I. Insert length ($Z^B$) & 634 & 399 \\
  II. Minus strand hanging reads & 3211 & 2935 \\
  III. Plus strand hanging reads & 2790 & 2461 \\
  Overlap between I and (II and III) at ends $^*$ & 58 ($9\%$) & 47 ($12\%$) \\
  \hline
\end{tabular}
\caption{\label{tab:NA12878calls}  Number of calls made on NA12878 chromosome 20 by a scan using only the insert length statistic (I), only the minus strand hanging reads statistic (II), or only the plus strand hanging reads statistic (III).  Results for two different FWER thresholds, $\alpha=10\%, 1\%$ are shown.  Row 4shows the number (and percentage) of regions from I that overlap with a region in II at the left end, and a region in III at the right end.}\end{center}\end{table}

\begin{figure}
\begin{center}
\includegraphics[scale=.45]{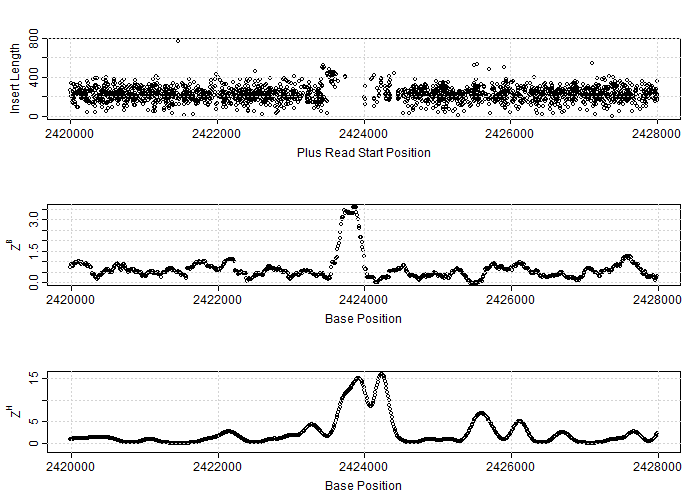} \\
\end{center}
\caption{\label{fig:data3} A region containing a putative homozygous deletion of roughly 200 base pairs on chromosome 20. Top plot shows the mapped insert length versus the start position of the plus strand read.  Middle plot shows the insert length score in this region.  Bottom plot shows $Z^H$, the hanging reads score.}
\end{figure}

\begin{figure}
\begin{center}
\includegraphics[scale=.45]{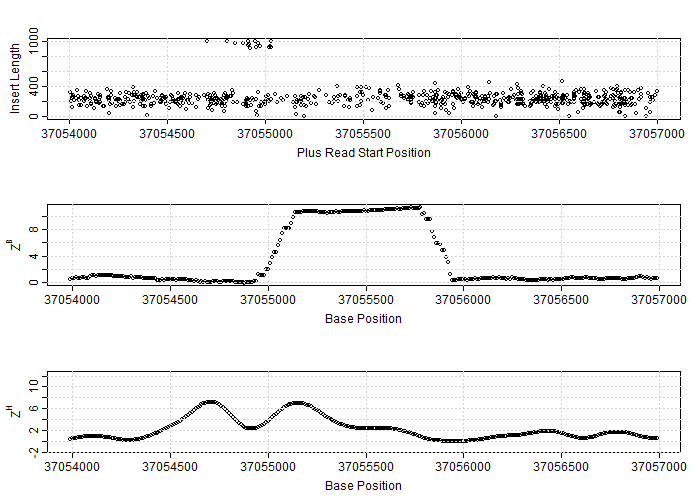} \\
\end{center}
\caption{\label{fig:data4}A region containing a putative heterozygous deletion of roughly 800 base pairs on chromosome 20.  Top plot shows the mapped insert length versus the start position of the plus strand read.  Middle plot shows the insert length score in this region.  Bottom plot shows $Z^H$, the hanging reads score.}
\end{figure}

Figures \ref{fig:data3} and \ref{fig:data4} show two example regions where either one or both scores have passed the threshold.  In Figure \ref{fig:data3}, there is a putative homozygous deletion of about 200 bases which generates the ideal pattern of a cluster of read pairs with shifted insert length preceding a region of no coverage (top plot).  Supporting this deletion are peaks in both $Z^+$ and $Z^-$. (The bottom plot shows their sum.)

Figure \ref{fig:data4} shows a cluster of read pairs with roughly 800 bp shift in insert length preceding a region of approximately 800 base pairs with coverage reduced by about half.  This visibly obvious tell tale pattern for a heterozygous deletion is convincing even without careful mathematical modeling.  Yet, at the $\alpha=10\%$ level there are no significant peaks in either the plus or minus strand hanging read scores.  Cases like this are not uncommon in the data.  There are also cases where strong evidence from the hanging reads are not supported by evidence from $Z^B$.  Real data is erratic, and ideal patterns like the example in Figure \ref{fig:data3} are the exception rather than the rule.

The genome scan that produces a list of candidate regions is only one of many steps in the analysis of such a rich data set.  To improve confidence and accuracy, regions such as those in Figures \ref{fig:data3} and \ref{fig:data4} should be analyzed more carefully, ideally by a more laborious local assembly of the reads that map to the region.

\section{Summary and Discussion} \label{sec:discussion}

We studied scan statistics for Poisson-type data, with
emphasis on several statistics that are useful for
detecting local genomic signals in next-generation sequencing experiments,
such as peak detection in ChIP-Seq and structural variant detection
by paired-end whole genome sequencing.
Despite their different formulations, analytic significance approximations
for these statistics can be obtained through a general framework that involves
embedding the statistics into an exponential family (\ref{lambda1})
and applying the measure transformation technique described in
Siegmund, Yakir and Zhang (2011).  See also
Yakir (2013).

Some of our analyses have been focused on a mixture model, which we
characterized using the kernel function (\ref{mixturekernel}).  This model
can be viewed as a simplified version of the model for bracketing
read pairs (\ref{del2}).  (A simplified version of the hanging reads
model is suggested in Remark 2 at the end of Section 3.3.)
We described significance level approximations
under this model in detail, and showed by
Monte Carlo that they are reasonably accurate.  We also conducted
power studies under this model, which reveal a complex picture regarding how
power depends on the choice of scanning parameter(s), the assumed homogeneity
of the process, and the values of nuisance parameters.  The key observations
are summarized in Section \ref{sec:power}.  We expect these qualitative
statements regarding power to generalize to the more complex models in
Section \ref{sec:bettermodel}, although some aspects neglected by
the simplified models, e.g., the asymmetry between insertions and
deletions,  may fail to be accurately illustrated.
For a numerical example, for parameters
associated with  the second row of
Table \ref{table:II}, where the bracketing pairs statistic for our better model to
detect insertions has marginal power 0.75, the simple mixture model
would have power 0.93;  the simplified model for hanging reads would
have marginal power 0.85
compared to 0.95 for the better model.

For structural variant detection using paired-end sequencing, we formulated
a model that incorporates three different features of the data:  Read
coverage, mapped insert length, and hanging read pairs.  The log likelihood
ratio scan statistic under this model is a sum of terms, which we
call ``scores,'' for each of these three features.
While the bracketing pairs statistics have increasing power to
detect longer deletions, their power to detect insertions first increases,
then decreases with the length of the insertion.  The power of the hanging
read statistics to detect deletions does not depend on the length of
the deletion, while their power to detect insertions increases with
the length of the insertion and approach an asymptote typically
less than one.

Although read coverage is suggested as one source of information in our model,
we have neglected it in the power calculations that we report in detail.
The reason is that the read coverage statistic usually performs poorly
unless the true value of $r$ is close to .5 and the deletion (for example) is
fairly long; in which case the statistic using mapped insert length is itself
reasonably powerful.  If we use the sum of $Z^C$ and $Z^B$ to call deletions, for most alternatives we have smaller marginal power compared to using $Z^B$ alone, although in specific cases the
marginal power does increase.  For
example, for rows 11-13 of Table \ref{table:DD}, the marginal power would become
0.38, 0.91, and 0.43 respectively.  For $r = 0.4$ or 0.5 and $w = 100$
or 150, the marginal
power of the sum $Z^C+Z^B$ is greater than that for $Z^B$ alone, but then the
marginal power of the hanging read statistic in these cases is even greater.

In the empirical data that we have examined, larger mean insert lengths also have
substantially larger standard deviations.  Also, such libraries tend to exhibit skewness and sometimes even multimodality.  A consequence
of the contemporary  move to increase read and insert length
is that relatively speaking the hanging read statistics gain power, but
the bracketing pairs statistics can lose substantial power.

Our analyses in Section \ref{sec:sv2} assume constant read coverage $\kappa$.  As mentioned in
Section \ref{sec:bettermodel}, mean read coverage has been empirically observed to fluctuate
along the genome and correlates with known features such as GC content.  Since the null
distribution of the scores depend on $\kappa$, if we allow $\kappa$ to vary the thresholds for the
scores would change with genome position.  Even with the analytic approximations for the p-values,
back-solving them to obtain appropriate thresholds for a given significance level requires
a substantially increased amount oft computation.  One could also conduct the scan using the
corresponding likelihood ratios $\ell_{\tau}=\beta Z_{\tau}-\psi(\beta)$, the thresholds for which
are much less variable as a function of $\kappa$.  However, the parameter $\beta$ in the likelihood ratio statistic would vary with
$\kappa$ and thus the computational issue can not be avoided.
A viable option in practice is to first segment the genome into blocks of approximately
homogeneous read coverage, then scan each block separately with a threshold computed using
the block-specific $\kappa$.  The global p-value would then be simply the sum of
the block-wise p-values.  In implementing this approach, one may want to ignore genomic
regions of low coverage.  One of the lessons of the simple mixture model is that a substantial
amount of power is inevitably lost in regions of low coverage and cannot be recovered
by a simple adjustment of the significance threshold.

An open question is how different statistics should be combined to improve detection
accuracy.  We found in our power analysis that summing the scores, as
in the log-likelihood, is rarely better than applying each score
individually.   The reason is that for most alternative settings there is
one score that dominates the others, and incorporating the others
by simple addition contributes mainly noise.  Thus, it may be better to apply each score individually
and then combine detections using a Bonferroni correction.  It may also be
better to combine the scores in a weighted sum, as in Senbaobaglu, Li and Zhang (2011).

While we have been focusing mainly on control of the family-wise error rate,
in genomic studies the false discovery rate (FDR) is often an appealing mode
of multiple testing control.  The boundary crossing probabilities can be
easily converted into the expected number of false discoveries under the null,
and used for FDR control as described in Siegmund, Yakir, and Zhang (2011).

\begin{center} {REFERENCES} \end{center}
\begin{description}

\item Abyzov, A., Urban, A., Snyder, M. and Gerstein, M. (2011). CNVnator: An approach to discover, genotype, and characterize typical and atypical CNVs from family and population genome sequencing.  {\sl Genome Research} {\bf 21}, 974-984.

\item Benjamini, Y. and Speed, T.P. (2012). Summarizing and correcting the GC content bias in high-throughput sequencing.  {\sl Nucleic Acids Research} {\bf 40}, e72.

\item Campbell, P. J., Stephens, P. J., Pleasance, E. D., O'Meara, S., Li, H., San-
tarius, T., Stebbings, L. A., Leroy, C., Edkins, S., Hardy, C., Teague, J. W.,
Menzies, A., Goodhead, I., Turner, D. J., Clee, C. M., Quail, M. A., Cox, A.,
Brown, C., Durbin, R., Hurles, M. E., Edwards, P. A. W., Bignell, G. R.,
Stratton, M. R. and Futreal, P. A. (2008). Identification of somatically acquired
rearrangements in cancer using genome-wide massively parallel paired-end sequencing.
{\sl Nature Genetics} {\bf 40}, 722-729.

\item Chan, H.P. and Zhang, N. R. (2006). Scan statistics with weighted observations. {\sl J. of American Statistics Association} {\bf 102}, 595-602.

\item Chiang, D. Y., Getz, G., Jaffe, D. B., O'Kelly, M. J., Zhao, X., Carter, S. L.,
Russ, C., Nusbaum, C., Meyerson, M. and Lander, E. S. (2009). High-resolution
mapping of copy-number alterations with massively parallel sequencing. {\sl Nature methods}
{\bf 6}, 99-103.

\item Feingold, E., Brown, P. O., and Siegmund, D. (1993).
Gaussian models for genetic linkage analysis using complete
high resolution maps of identity by descent.
{\sl Am. J. Hum. Genet.} {\bf 53},  234-251.

\item Karlin, S., Dembo, A. and Kawabata, K. (1990). Statistical composition of high-scoring segments from molecular sequences. {\sl Annals of Statistics} {\bf 18}, 571-581.

\item Lander, E.S. and Botstein, D. (1989).
Mapping mendelian factors underlying quantitative traits using RFLP linkage maps.  {\sl Genetics} {\bf 121}, 185-199.

\item Medvedev P, Stanciu M, Brudno M. (2009) Computational methods for discovering structural variation with next generation sequencing. {\sl Nature Methods} {\bf 6}, S13-20.

\item Peng, Jie and Siegmund, D. (2005).
The admixture model in linkage analysis,
{\sl J. Statist. Planning and Inference} {\bf 130}, 317-324.

\item Rabinowitz, D. and Siegmund, D. (1997).
The approximate distribution of the maximum of a smoothed
Poisson random field,
{\sl Statistica Sinica} (1997) {\bf 7} 167-180.

\item  Schwartzman, A.,  Jaffe, A.,  Gavrilov, Y. and  Meyer, C.E.
(2013).  Multiple testing of local maxima for detection of peaks in ChIP-seq
data, {\sl Annals of Applied Statistics}, {\bf 7}, 471-494.

\item Shen, J. and Zhang, N. R. (2012).  Change-point model on nonhomogeneous Poisson processes with application in copy number profiling by next-generation DNA sequencing. {\sl Annals of Applied Statistics} {\bf 6}, 476-496.

\item Siegmund, D. and Worsley, K. (1995).
Testing for a signal with unknown location and scale in a
stationary Gaussian random field, {\sl Ann. Statist.} {\bf 23} 608-639.

\item Siegmund, D., Yakir, B. and Zhang, N.R. (2010).
Tail approximations for maxima of random fields by
likelihood ratio transformations,
{\sl Sequential Analysis}. {\bf 29}, 245-262.

\item Siegmund, D., Yakir, B. and Zhang, N.R. (2011).
Detectiing simultaneous variant intervals in aligned
sequences,
{\sl Ann. Appl. Statist.} {\bf 5} 645-668.

\item Siegmund, D., Yakir, B. and Zhang, N.R. (2011).
False discovery rate for scan statistics,
{\sl Biometrika} {\bf 98}, 979-986.

\item Tang, H.K. and Siegmund, D. (2001).
Mapping quantitative trait loci in oligogenic models,
{\sl Biostatistics} {\bf 2}, 147-162.

\item
Worsley, K.J., Evans, A.C., Marrett, S. and Neelin, P. (1992).
A three dimensional statistical analysis for CBF activation studies
in human brain, {\sl Journal of Cerebral Blood Flow and Metabolism}
{\bf 12}, 900-918.

\item Yakir, B. (2013). {\sl Extremes in Random Fields:  A Theory and Its
Applications}, Wiley, Chichester,
United Kingdom.

\end{description}

\end{document}